\documentclass[aps,prd,twocolumn,superscriptaddress,showpacs,floatfix,handout,10pt]{revtex4-1}

\usepackage{CJKutf8} 

\usepackage{amsmath,amssymb,mathtools}
\usepackage{color}
\usepackage[table]{xcolor}
\usepackage{graphicx}
\usepackage{subfigure}
\usepackage[colorlinks=true,linkcolor=blue, citecolor=red, urlcolor=blue, bookmarks]{hyperref}
\usepackage{multirow,makecell} 
\usepackage{textcomp}
\usepackage{wasysym}
\usepackage{ulem}
\usepackage{tikz}
\usetikzlibrary{shapes,positioning}
\usepackage{epstopdf}
\usepackage{graphicx}
\usepackage{subcaption}

\newcommand{\RNum}[1]{\uppercase\expandafter{\romannumeral #1\relax}}

\def \be {\begin{equation}}
\def \ee {\end{equation}}
\def \ba {\begin{array}}
\def \ea {\end{array}}
\def \bea {\begin{eqnarray}}
\def \eea {\end{eqnarray}}

\def \ble {\begin{widetext}\begin{equation}}
\def \ele {\end{equation}\end{widetext}}
\def \blea {\begin{widetext}\begin{eqnarray}}
\def \elea {\end{eqnarray}\end{widetext}}

\def \tr {\mathrm{tr}}

\def \and {{\mathrm{and}}}

\newcommand{\bra}[1]{\ensuremath{\left\langle#1\right|}}
\newcommand{\ket}[1]{\ensuremath{\left|#1\right\rangle}}

\newtheorem{theorem}{Theorem}

\begin{document}

\begin{CJK*}{UTF8}{gbsn}

\title{Explicit Pfaffian Formula for Amplitudes of Fermionic Gaussian Pure
States in Arbitrary Pauli Bases}

\author{M.~A.~Rajabpour }

\affiliation{Instituto de Fisica, Universidade Federal Fluminense,
Av.~Gal.~Milton Tavares de Souza s/n, Gragoat\'a, 24210-346, Niter\'oi, RJ, Brazil}

\author{ M. A. Seifi Mirjafarlou}

\affiliation{Instituto de Fisica, Universidade Federal Fluminense,
Av.~Gal.~Milton Tavares de Souza s/n, Gragoat\'a, 24210-346, Niter\'oi, RJ, Brazil}

\author{Reyhaneh Khasseh}
\affiliation{Theoretical Physics III, Center for Electronic Correlations and Magnetism,
Institute of Physics, University of Augsburg, D-86135 Augsburg, Germany}

\begin{abstract}
The explicit computation of amplitudes for fermionic Gaussian pure states in arbitrary Pauli bases is a long-standing challenge in quantum many-body physics, with significant implications for quantum tomography, experimental studies, and quantum dynamics. These calculations are essential for analyzing complex properties beyond traditional measures, such as formation probabilities, global entanglement, and entropy in non-standard bases, where exact and computationally efficient methods remain underdeveloped.
In addition to these physical applications, having explicit formulas is crucial for optimizing negative log-likelihood functions in quantum tomography, a key task in the NISQ era.
In this work, we present an explicit Pfaffian formula (Theorem \ref{theorem1}) for determining these amplitudes in arbitrary Pauli bases, utilizing a matrix whose structure reflects the qubit parity. Additionally, we introduce a recursive relation (Theorem \ref{theorem2recursiveformula}) that connects amplitudes for systems with varying qubit numbers, enabling scalable computations for large systems.
Together, these results provide a versatile framework for studying global entanglement, Shannon-Rényi entropies, formation probabilities, and performing efficient quantum tomography, thereby significantly expanding the computational toolkit for analyzing complex quantum systems. 
Finally, we utilize our formalism to determine the post-measurement entanglement entropy, reflecting how local measurements alter entanglement, and compare the outcomes with conformal field theory predictions.
\end{abstract}

\maketitle

\end{CJK*}

\section{Introduction}

Fermionic Gaussian states are among the earliest and most widely studied many-body fermionic systems, with applications spanning a range of fields. They serve as exact ground states for coupled spin chains \cite{LIEB1961407}, provide approximate ground states for interacting fermions \cite{Thouless1960,szabo2012modern,EA2007,Kraus2010,Terhal2023}, and play an important role in impurity problems \cite{Bravyi2017}. Beyond these, they are valuable in quantum simulation, benchmarking quantum computation \cite{Knill2001,Terhal2002,Brod2016}, and modeling the time evolution of free fermionic systems after a quantum quench \cite{CEF2011}. While several properties of these states are well-understood—often traceable back to Wick’s theorem, such as entanglement entropy \cite{Peschel2001,Kitaev2003} and fidelity \cite{Zanardi2008}—other characteristics, like efficient tomography \cite{Gluza2018,Gorman2022,Leone2024}, trace distance \cite{RZ2023,Wang2024,Leone2024}, and Shannon-R\'enyi entropy \cite{NRR2024}, have sparked intense recent research. It is evident, however, that calculating all relevant quantities for Gaussian states does not always simplify to Wick’s theorem alone, leading some aspects to be categorized as challenging problems. For an overview of both analytical and numerical techniques, we refer the reader to Ref.~\cite{Surace2022}, and references therein.

A central yet unresolved property of Gaussian states is the explicit expression of their amplitudes in the arbitrary Pauli local basis—a spin representation achieved via the Jordan-Wigner transformation. When fermionic Gaussian pure states are expressed in the computational basis, their amplitudes correspond to the Pfaffians of specific submatrices (or "Pfaffinhos") of an antisymmetric matrix \cite{Bravyi2005,Becca-Sorella2017,TKR2024}. This structure enables the efficient computation of all amplitudes through algorithms optimized for Pfaffian calculation. However, in spin systems, alternative bases often provide more pertinent information. Determining the amplitudes in such alternative bases is particularly valuable in the following contexts:

The primary application of an explicit formula for Gaussian state amplitudes lies in efficiently calculating probabilities. With a polynomial-time evaluable formula, it becomes feasible to determine the probability of observing a specific bit string following a projective measurement in a chosen Pauli basis, even for large systems, i.e. $L>1000$. In quantum spin chains, this probability, known as the formation probability, has been explored in several studies, particularly in the $\sigma^z$ and $\sigma^x$ bases \cite{Essler1994,Shiroishi2001,Franchini2005,Stephan2009,NR2016}, with recent work extending the analysis to the $x-y$ plane \cite{TKR2024}.

Another immediate application is the calculation of global entanglement in fermionic Gaussian states \cite{Wei2005}. This measure is directly tied to the maximum formation probability, as defined above. Specifically, to compute global entanglement, one first determines the formation probabilities across all possible Pauli bases, identifying the highest among them. Global entanglement, as defined in \cite{Wei2003}, is then given by the negative logarithm of this maximum probability. It can also be viewed as a measure which its generalization can be viewed as multipartite entanglement \cite{Sen2010}. While calculating it typically requires optimization over all angles—a task that is often challenging—having an exact expression for any angle provides a significant advantage.

A third application arises in calculating Shannon-R\'enyi entropy in alternative bases, where the entropy depends on all possible probabilities. This entropy measure has been widely studied in quantum spin chains \cite{Stephan2009,Stephan2010,Alcaraz2013,Stephan:2014,Alcaraz2014,Tarighi2022} and was recently utilized to determine the central charge of critical quantum spin chains, which can be described by conformal field theory, in superconducting qubits \cite{central-charge}.
 Since the number of probabilities grows exponentially with the qubit count, calculating these values poses a significant computational challenge. However, an explicit formula enables more efficient analytical solutions or code optimizations, making it feasible to explore larger system sizes.

Another application is to examine the quantum dynamics of spin systems that are mappable to free fermions. For example, in a quantum quench scenario with the transverse field Ising chain, where the initial state is the ground state under a different transverse field, the system evolves as a Gaussian state over time\cite{CEF2011}. Using an explicit formula, transition amplitudes in alternative bases can be computed efficiently after projective measurements. The same principle applies if the system begins in a local product state in arbitrary local Pauli basis, with transition amplitudes to the local $\sigma^z$ basis computed through projective measurement in the computational basis. These transition amplitudes have already been investigated in the context of dynamical phase transitions \cite{Heyl2013,Heyl2018,NRV2020} and experimentally measured in the $\sigma^x$ basis \cite{Jurcevic2017}. However, systematically calculating these probabilities for arbitrary product states, whether as the initial state or as the basis for projective measurement, remained an open problem.

The fifth application concerns the tomography of Gaussian states, a field of growing interest due to its relevance in benchmarking quantum computation \cite{Gorman2022,Leone2024}. In practical scenarios, projective measurements in local Pauli bases are required experimentally to reconstruct the quantum state. Some lower bounds have been established for the number of copies needed to determine the state’s correlation matrix within a specific error margin \cite{Leone2024}.
For effective tomography of a Gaussian pure state, it is crucial to have explicit amplitude expressions as functions of the Gaussian state parameters, allowing for optimal parameter estimation based on experimentally obtained probabilities. For instance, one might conduct measurements in specific bases that are either physically relevant or experimentally feasible, followed by minimizing quantities like the negative log-likelihood \cite{Najafi2021}. To implement this in practical cases—such as the tomography of the ground state of the transverse field Ising models—an explicit form of Gaussian states in alternative bases is essential\cite{ZR2025}.

{\color{red}}

Another potential application lies in calculating post-measurement entanglement entropy \cite{R2015,Baweja2024} and measurement-induced entanglement \cite{Potter2024}, which refers to the entanglement entropy of a system following a projective measurement. For post-measurement entanglement entropy, a projective measurement is performed in a local basis on part of the system, followed by a calculation of the bipartite entanglement entropy of the remaining system. In the case of measurement-induced entanglement, one averages overall possible entanglement values following projective measurement. Having explicit formulas for probability amplitudes in alternative bases could enhance numerical algorithms for computing these quantities in free fermionic systems, such as the Ising and XY chains.


Despite their wide applicability, fermionic Gaussian states constitute a restricted subset of quantum states. They are efficiently classically simulable, much like stabilizer states in spin systems, and typically fail to capture non-Gaussian correlations. These limitations mean that Gaussian states are not suitable for describing certain interacting quantum phases or for achieving computational advantage in the general case. Accordingly, our framework is confined to the domain of free or weakly interacting systems, where Gaussian descriptions remain powerful. In this context, our explicit methods provide valuable tools for state characterization, efficient simulation, and benchmarking. Moreover, the insights and structural results presented here may inform future extensions toward non-Gaussian ansätze or hybrid methods capable of addressing more complex quantum regimes.

Within this setting, it is known that certain classes of fermionic systems—such as those described by match gates—can be simulated efficiently on a classical computer \cite{Terhal2002,Brod2016}. This suggests that explicit, closed-form expressions for measurement amplitudes should exist, at least in principle.
However, existing methods do not appear to provide a straightforward approach to derive an explicit formula for the amplitudes. In this paper, we present an explicit formula for the amplitudes of fermionic Gaussian pure states in an arbitrary local Pauli basis. This formula is expressed in terms of the Pfaffian of a matrix with dimensions \( L \) or \( L+1 \), depending on whether the system has an even or odd number of qubits, respectively, enabling efficient computation of individual amplitudes even for large systems (\( L > 1000 \)). We then introduce a recursive formula that relates the amplitudes of an \( L \)-qubit system to those of systems with \( L-2 \) and \( 2 \) qubits. In non-computational bases, this relationship is not necessarily within the same basis; for example, the amplitude configurations in the \( x \)-basis can be related to those in the \( y \)-basis.

Consider a qubit state expressed in the computational basis. To transform this state into a new, generic basis, the following unitary matrix can be applied:
\begin{equation}{\label{Generic U matrix}}
    \bold{U}_{(\phi,\theta,\alpha)}=\left(
\begin{array}{ccccccc}
    \cos{\frac{\theta}{2}}    & \sin{\frac{\theta}{2}}e^{-i\phi}  \\
 \sin{\frac{\theta}{2}}e^{-i\alpha}  &  -\cos{\frac{\theta}{2}}e^{-i(\alpha+\phi)}   \\
\end{array}
\right).
\end{equation}
For example, $(\phi,\theta,\alpha)=(0,\frac{\pi}{2},0)$ and $(\frac{\pi}{2},\frac{\pi}{2},0)$ are associated to the $\sigma^x$ and $\sigma^y$ bases respectively.

To rewrite a state $\ket{\psi}_z$ expressed in computational basis in the $(\boldsymbol{\phi},\boldsymbol{\theta},\boldsymbol{\alpha})$ basis one may  use the mapping
\begin{widetext}

\begin{equation}{\label{U matrix base change}}
    \ket{\psi}_{(\boldsymbol{\phi},\boldsymbol{\theta},\boldsymbol{\alpha})} =\bold{U}_{(\boldsymbol{\phi},\boldsymbol{\theta},\boldsymbol{\alpha})}\ket{\psi}_z= \bold{U}_{(\phi_1,\theta_1,\alpha_1)}\otimes \bold{U}_{(\phi_2,\theta_2,\alpha_2)}...\otimes \bold{U}_{(\phi_L,\theta_L,\alpha_L)} \ket{\psi}_z.
\end{equation}
\end{widetext}
In this paper, we assume that the state $\ket{\psi}_z$ is a fermionic Gaussian state.

We start the section II with an overview of the properties of free fermion Gaussian pure states, formulated in a way that facilitates the derivation of our results. In section III, we proceed by performing a basis transformation to express the amplitudes as an exponential sum. This sum is then simplified to yield a single explicit Pfaffian formula. In section IV, we write an explicit formula for probabilities that can be used for many applications. Building on this, section V shows how our framework allows for the efficient computation of post-measurement entanglement entropy in arbitrary Pauli bases. Finally, we conclude the paper in section VI by mentioning some possible future generalizations and applications.
\section{Gaussian pure states}
In reference \cite{TKR2024}, it was demonstrated that any arbitrary Gaussian pure state can be transformed into the following form:
\begin{equation}{\label{GPS-general}}
    \ket{\bold{R},\mathcal{C}} = \frac{1}{\mathcal{N}_R}e^{{\frac{1}{2}\sum_{i,j}^La_{i}r_{ij}a_{j}}}\ket{\mathcal{C}},
\end{equation}
where  $a_j=c_j(c_j^{\dagger})$ if there is (not) fermion at site $j$ of the configuration $\mathcal{C}$ and  $\mathcal{N}_R=\det(\bold{I}+\bold{R}^{\dagger}.\bold{R})^{\frac{1}{4}}$. Note that, without loss of generality, we can assume that the \( \bold{R} \) matrix is antisymmetric. In summary, a Gaussian pure state is characterized by the \( \bold{R} \) matrix and the base configuration \( \mathcal{C} \).
To express the above state in the computational basis, we begin by defining ${\ket{\mathcal{C}}}={\ket{n_1, n_2, \ldots, n_L}}$ and ${\ket{\mathcal{I}}}={\ket{m_1, m_2, \ldots, m_L}}$, where $n_j, m_j \in \{0,1\}$. Then,  one can write \cite{TKR2024}:

\begin{equation}{\label{GPS-general-pfaffinho}}
    \ket{\bold{R},\mathcal{C}} = \frac{1}{\mathcal{N}_R}\sum_{\mathcal{I}}\text{sgn}(\mathcal{C},\mathcal{I})\bold{pf}\ \bold{R}_{\mathcal{I}}(\mathcal{C}){\ket{\mathcal{I}}},
\end{equation}
where $\bold{R}_{\mathcal{I}}(\mathcal{C})$ is the submatrix of $\bold{R}$ obtained by retaining only rows and columns $j\in\{1,2,...,L\}$ for which $|n_j - m_j| = 1$. The sign is given by $\text{sgn}(\mathcal{C},\mathcal{I}) = \prod_{i=2}^L (-1)^{|n_i - m_i| \sum_{j < i} n_j}$.

When the amplitude of the configuration $\ket{\mathcal{I}}=\ket{\bold{0}}$ is not zero one can rewrite the state as $\ket{\bold{R},\mathcal{C}} =  \ket{\bold{R}',\bold{0}}$.
The elements of the matrix $\bold{R}'$, i.e. $r'_{ij}$, can be obtained following the procedure explained in the Appendix \ref{sec:baseconfigurationchange}.
Consequently, the state in the configuration basis ends up having the following simple form
\begin{equation}{\label{GPS0-pfaffinho}}
  \ket{\bold{R},\mathcal{C}} =  \ket{\bold{R}',\bold{0}} = \frac{1}{\mathcal{N}_{R'}}\sum_{\mathcal{I}}\bold{pf}\ \bold{R}'_{\mathcal{I}}{\ket{\mathcal{I}}},
\end{equation}
where $\mathcal{I}$ is the bit string configuration,  $\bold{R}'_{\mathcal{I}}$ is the submatrix of the matrix $\bold{R}'$ in which we removed the rows and columns corresponding to the sites that there is no fermion. From this point onward, we will omit the prime notation from the \( \mathbf{R} \) matrix and take ${\ket{\mathcal{C}}}={\ket{\boldsymbol{0}}}$.

\section{Gaussian states in alternative bases}

Consider the following mapping called Jordan-Wigner (JW) transformation:

\begin{equation}{\label{JW}}
    c_l = \prod_{j<l}(-\sigma^{z}_j)\sigma^{-}_l, 
    \hspace{0.5cm}
    c^{\dagger}_l = \prod_{j<l}(-\sigma^{z}_j)\sigma^{+}_l.
\end{equation}
It is easy to show that $\sigma_l^z=2c_l^{\dagger}c_l-1$ which means that one can write the Gaussian state in the $\sigma^z$ basis by just the substitutions $\ket{0}\rightarrow \ket{\downarrow}_z$ and $\ket{1}\rightarrow \ket{\uparrow}_z$. The next formula provides our main theorem which is an explicit formula for the amplitudes.

\begin{widetext}
\begin{theorem}[Fermionic Gaussian Pure State in an arbitrary Pauli Basis]\label{theorem1}
Consider a Gaussian pure state $\ket{\bold{R},\bold{0}}$ with $L$ qubits. The amplitude of $\ket{\mathcal{S}}=\ket{s_1,s_2,...,s_L}$ where $s_j$ is in the basis $(\phi_j,\theta_j,\alpha_j)$ can be written as:
\begin{eqnarray}{\label{amplitude-arbitrary-basis}}
 a_{\mathcal{S}}(\bold{R},\boldsymbol{\phi},\boldsymbol{\theta},\boldsymbol{\alpha})=\frac{(-1)^{L(1-\bar{s}_1)/2}\sqrt{2}^{L\text{mod} \ 2}}{\mathcal{N}_{R}}e^{-i\sum_{j\in\mathcal{S}^-}\alpha_j}\prod_{j\in \mathcal{S}^+}\cos\frac{\theta_j}{2}\prod_{j\in \mathcal{S}^-}\sin\frac{\theta_j}{2}\bold{pf} \bold{R}^{\mathcal{S}}.
\end{eqnarray}
When $L$ is even
\begin{eqnarray}\label{RS-matrix}
 R_{nm}^{\mathcal{S}}(\boldsymbol{\phi},\boldsymbol{\theta},\boldsymbol{\alpha})= 
 r_{nm}e^{i(\phi_n+\phi_m)}+(-1)^{n+m}(-1)^{\frac{\bar{s}_n+\bar{s}_m}{2}}\tan^{\bar{s}_n}{\frac{\theta_n}{2}}\tan^{\bar{s}_m}\frac{\theta_m}{2}
\end{eqnarray}
and $\mathcal{S^+}$ and $\mathcal{S^-}$ are the set of qubits with spins in the direction of up or down respectively and we associated $\bar{s}=\pm1$ to spins $up \ (down)$. When $L$ is odd we first add a zero row and column to the matrix $\bold{R}$ and use the same formula as above with the conditions $\bar{s}_{L+1}=\bar{s}_{1}$ and $\theta_{L+1}=\frac{\pi}{2}$ and $\alpha_{L+1}=0$.
\end{theorem}
\end{widetext}

The proof of the theorem above is provided in the Appendix \ref{sec:theorem1}. It begins by identifying the coefficients of the trigonometric functions, which are composed of pfaffinhos, and then reformulates these trigonometric terms into a single Pfaffian expression. This approach is closely related to a generalization \cite{Lieb2016} of Lieb’s theorem \cite{Lieb1968} which we further extend it to odd values of $L$ in the Appendix \ref{sec:theorem1}.  It is worthwhile to discuss a few important points: Firstly, the formula remains valid even when \( \theta_j \to 0 \). 
 In Appendix \ref{S3} we provide an alternative formula in which the contribution of the angle $\theta$ directly appears as positive powers of \( \cos\left(\frac{\theta}{2}\right) \) and \( \sin\left(\frac{\theta}{2}\right) \) which makes the above point clear.
Additionally, if the sign of all spins is reversed, then, apart from a phase factor, the amplitude of \( \ket{-\mathcal{S}} \) can be obtained by substituting \( \theta_j \to \pi - \theta_j \). When $\theta_j=\frac{\pi}{2}$ apart from a phase changing the sign of all the spins $\bar{s}_j$ does not change the amplitude. This is partially the reason for the success of the domain wall representation in the ref \cite{TKR2024} for this special case. 

We also have the following identity
\begin{eqnarray}{\label{R-minus-R}}
a_{\mathcal{S}}(\bold{R},\boldsymbol{\phi},\boldsymbol{\theta},\boldsymbol{\alpha})=a_{\mathcal{S}}(-\bold{R},\boldsymbol{\phi}+\frac{\pi}{2},\boldsymbol{\theta},\boldsymbol{\alpha}).
\end{eqnarray}
Since the Gaussian pure states are characterized by the $\frac{L(L-1)}{2}$ elements of the matrix $\mathbf{R}$, one would expect to have the same number of independent amplitudes. For example, when $\theta \neq \frac{\pi}{2}$, all amplitudes can be expressed in terms of the amplitudes corresponding to configurations where only two spins are in the positive direction. Examples of such cases are provided in the Appendix \ref{Relationsamongamplitudes}.

The Pfaffian form of the amplitudes offers an intriguing approach to derive amplitudes for systems with larger qubit numbers by building on those with fewer qubits. This approach resembles Wick’s theorem for correlation functions; however, it is important to note that this is not a straightforward translation to the standard Wick's theorem. In particular, correlation functions of operators that are non-local with respect to the fermionic representation do not adhere to the standard Wick’s formula. Additionally, in an arbitrary Pauli basis, the amplitudes for systems with an odd number of qubits also depend on the amplitudes of individual qubits. The following theorem is our second main result.

\begin{widetext}
\begin{theorem}[A recursive theorem for amplitudes]\label{theorem2recursiveformula}

Consider a Gaussian pure state $\ket{\bold{R},\bold{0}}$ with $L$ qubits and $P(D)=\prod\limits_{k\in D}\bar{s}_k$ as the parity of spins in the region $D$. The amplitude of $\ket{\mathcal{S}}=\ket{s_1,s_2,...,s_L}$ where $s_j$ is in the basis $(\phi_j,\theta_j,\alpha_j)$ can be written as:
\begin{eqnarray}{\label{theorem2}}
 b_{\mathcal{S}}(\boldsymbol{\phi},\boldsymbol{\theta},\boldsymbol{\alpha})=\hspace{10cm}\nonumber\\\sum_{j=1}^{L'/2} P(\bar{A}_{2j})b_{s_1,s_{2j}}(\boldsymbol{\phi},\boldsymbol{\theta},\boldsymbol{\alpha})b_{\mathcal{S}/\{s_1,s_{2j}\}}(\boldsymbol{\phi},\bar{\boldsymbol{\theta}}^j,\boldsymbol{\alpha})
 -\sum_{j=1}^{L'/2-1} P(A_{2j})b_{s_1,s_{2j+1}}(\boldsymbol{\phi}+\frac{\pi}{2},\boldsymbol{\theta},\boldsymbol{\alpha})b_{\mathcal{S}/\{s_1,s_{2j+1}\}}(\boldsymbol{\phi},2\pi-\bar{\boldsymbol{\theta}}^j,\boldsymbol{\alpha}),
\end{eqnarray}
where $b_{\mathcal{S}}(\boldsymbol{\phi},\boldsymbol{\theta},\boldsymbol{\alpha})=\mathcal{N}_{R} \ a_{\mathcal{S}}(\boldsymbol{\phi},\boldsymbol{\theta},\boldsymbol{\alpha})$, $b_{s_n,s_{m}}=\mathcal{N}_{R_{nm}}a_{s_n,s_{m}}$ and $b_{\mathcal{S}/\{s_n,s_{m}\}}=\mathcal{N}_{R_{\overline{nm}}}a_{\mathcal{S}/\{s_n,s_{m}\}}$ in which 
$\overline{nm}$ is the complement of the $nm$. $\bold{R}_{nm}$ is the matrix $\bold{R}$ in which we keep the rows and columns $n,m$ and $\bold{R}_{\overline{nm}}$ is the matrix $\bold{R}$ in which we remove the rows and columns $n,m$. The normalization $\mathcal{N}_{R_{nm}}=\det\left(\bold{I}+\bold{R}^{\dagger}_{nm}.\bold{R}_{nm}\right)^{\frac{1}{4}}$ and $\mathcal{N}_{{R}_{\overline{nm}}}=\det\left(\bold{I}+\bold{R}_{\overline{nm}}^{\dagger}.\bold{R}_{\overline{nm}}\right)^{\frac{1}{4}}$. $L'=L[L+1]$ for even[odd] sizes and $A_{2j}=\{2,3,...,2j\}$, $\bar{A}_{2j}=\{2j+1,2j+2,...,L'\}$ and $\bar{\boldsymbol{\theta}}^j=\begin{cases}
    2\pi-\theta_k\hspace{0.5cm}k\in\bar{A}_{2j}\\
    \theta_k\hspace{1.3cm}k\in A_{2j}
\end{cases}$. For odd sizes $\bar{s}_{L+1}=\bar{s}_1$ and $\theta_{L+1}=\frac{\pi}{2}$ and $\alpha_{L+1}=0$.
\end{theorem}
\end{widetext}
The above theorem follows directly from Theorem \ref{theorem1}, relying on fundamental properties of the Pfaffian; see Appendix \ref{Alternativeformulafortherecursiveformula} for the proof and a list of examples. This recursive formula is particularly useful for incremental amplitude calculation in large systems, as it allows amplitudes to be computed progressively as qubits are added, without the need to recompute the full Pfaffian at each step, provided the $\bold{R}$ matrix is known. Furthermore, when amplitudes for a smaller system are already available, the theorem facilitates the efficient calculation of amplitudes for a larger system without explicitly requiring the $\bold{R}$ matrix. Notably, the recursive structure is advantageous for simulating quantum circuits, such as matchgate circuits, enabling the incremental evolution of fermionic Gaussian states while reducing computational complexity. This approach also supports parallelization, making it a powerful tool for simulating quantum algorithms and benchmarking quantum processors.
Some other alternative forms for the amplitude relations are provided in the Appendix \ref{S3}. In particular, we provide two formulas that connect the amplitudes of $L$-qubit systems to those with just $2$ qubits but with adjusted angles.

\section{Formation Probabilities}
 A simple application of the theorem \ref{theorem1} leads to the following result for the formation probabilities:
\begin{widetext}
\begin{eqnarray}{\label{EFP}}
 P_{\mathcal{S}}(\boldsymbol{\phi},\boldsymbol{\theta},\boldsymbol{\alpha})=2^{L \text{mod} \ 2}\Big{(}\frac{\det \bold{R}^{\mathcal{S}\dagger}(\boldsymbol{\phi},\boldsymbol{\theta},\boldsymbol{\alpha}).\bold{R}^{\mathcal{S}}(\boldsymbol{\phi},\boldsymbol{\theta},\boldsymbol{\alpha})}{\det (\bold{I}+\bold{R}^{\dagger}.\bold{R})}\Big{)}^{\frac{1}{2}}\prod_{j\in \mathcal{S}^+}\cos^2\frac{\theta_j}{2}\prod_{j\in \mathcal{S}^-}\sin^2\frac{\theta_j}{2}.
\end{eqnarray}
\end{widetext}

Here, \(P_{\mathcal{S}}(\boldsymbol{\phi},\boldsymbol{\theta},\boldsymbol{\alpha})\) denotes the probability of obtaining the outcome \(\ket{\mathcal{S}} = \ket{s_1, s_2, \dots, s_L}\) after a projective measurement in the basis specified by \((\boldsymbol{\phi},\boldsymbol{\theta},\boldsymbol{\alpha})\). Obviously, the angle $\alpha$ does not play any role in the probability formula. The above formula can be directly applied to numerically or analytically compute formation probabilities, Shannon-R\'enyi entropies, geometric entropy, transition probabilities in quantum quenches, and to optimize the Negative Log-Likelihood (NLL) in quantum tomography\cite{ZR2025}.
\begin{figure}
\begin{tikzpicture}[thick, scale=1.2, every node/.style={scale=1.}]
  \fill[blue!10] (0,0) circle(2);

  \draw[black, thick] (0,0) circle(1.5);

  \draw[black, thick, fill=white] (45:1.3) arc (45:90:1.3) -- (90:1.7) arc (90:45:1.7) -- cycle;
  \node at (67.5:1.5) {\large $A_1$};

  \draw[black, thick, fill=white] (225:1.3) arc (225:270:1.3) -- (270:1.7) arc (270:225:1.7) -- cycle;
  \node at (247.5:1.5) {\large $A_2$};

  \node at (157.5:1.8) {\large $B_1$};
  \node at (337.5:1.75) {\large $B_2$};

\end{tikzpicture}
\caption{Schematic of a one-dimensional chain with periodic boundary conditions, forming a closed ring. The chain is divided into two regions, $A$ and $B$, where $A = A_1 \cup A_2$ and $B = B_1 \cup B_2$. A projective measurement is applied on the $B_1$ and $B_2$ subsystems. We then study how the entanglement between $A_1$ and $A_2$ depends on the distance separating them along the ring.}
\label{fig:setup}
\end{figure}
\section{Post measurement entanglement entropy}
Post-measurement entanglement entropy reveals how local measurements reshape quantum correlations within a system. While CFT provides predictions in specific cases~\cite{Rajabpour2015,Antonini2023,Numasawa2016,Guo2019,Antonini2022,Weinstein2023}, calculating this quantity in a general measurement basis is challenging. Our explicit formulas offer a practical and general method to compute the entanglement entropy after projective measurements in arbitrary Pauli bases, making it possible to extract scaling behavior and decay exponents.

To demonstrate this, consider a bipartition of the system into subsystems \(A\) and \(B\). We perform a local projective measurement on subsystem \(B\) in the basis defined by the angles \((\boldsymbol{\phi},\boldsymbol{\theta},\boldsymbol{\alpha})\). A schematic illustration of this setup is shown in Figure~\ref{fig:setup}. In our framework, the full state is expressed in terms of a configuration \(\mathcal{S}\) that decomposes as
\[
\mathcal{S}=(\mathcal{S}_A,\mathcal{S}_B),
\]
with the probability of the full configuration given by Eq.~\eqref{EFP}. Consequently, the probability of obtaining a specific measurement outcome \(\mathcal{S}_B\) in subsystem \(B\) is obtained by summing over all configurations in subsystem \(A\):
\begin{equation}
\label{PSB}
P_{\mathcal{S}_B}(\boldsymbol{\phi},\boldsymbol{\theta},\boldsymbol{\alpha})
=\sum_{\mathcal{S}_A} P_{\mathcal{S}}(\boldsymbol{\phi},\boldsymbol{\theta},\boldsymbol{\alpha})\,.
\end{equation}
\begin{figure}
    \includegraphics[width=\columnwidth]{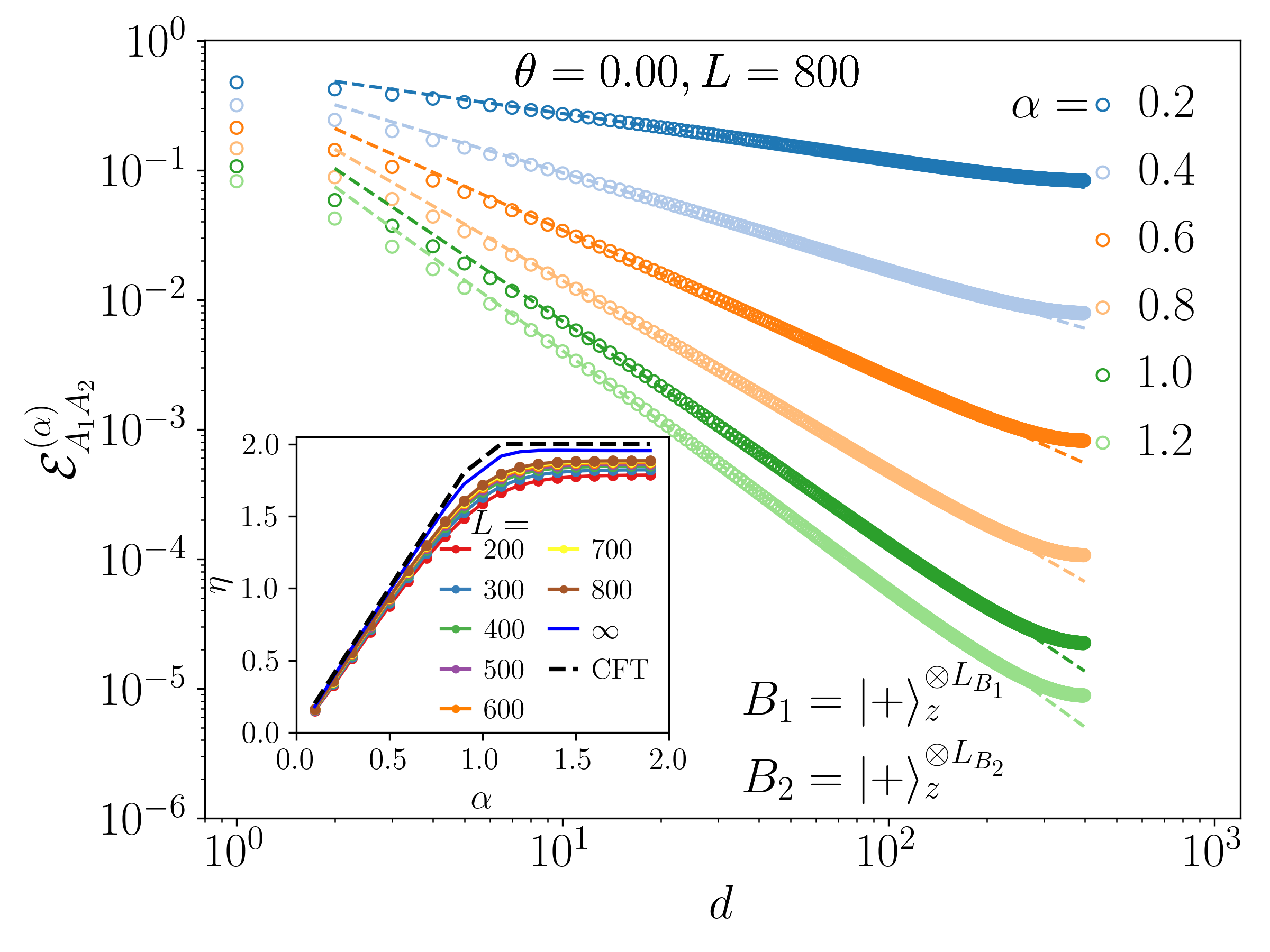}\put(-250,160){\textbf{(a)}}\\
\includegraphics[width=\columnwidth]{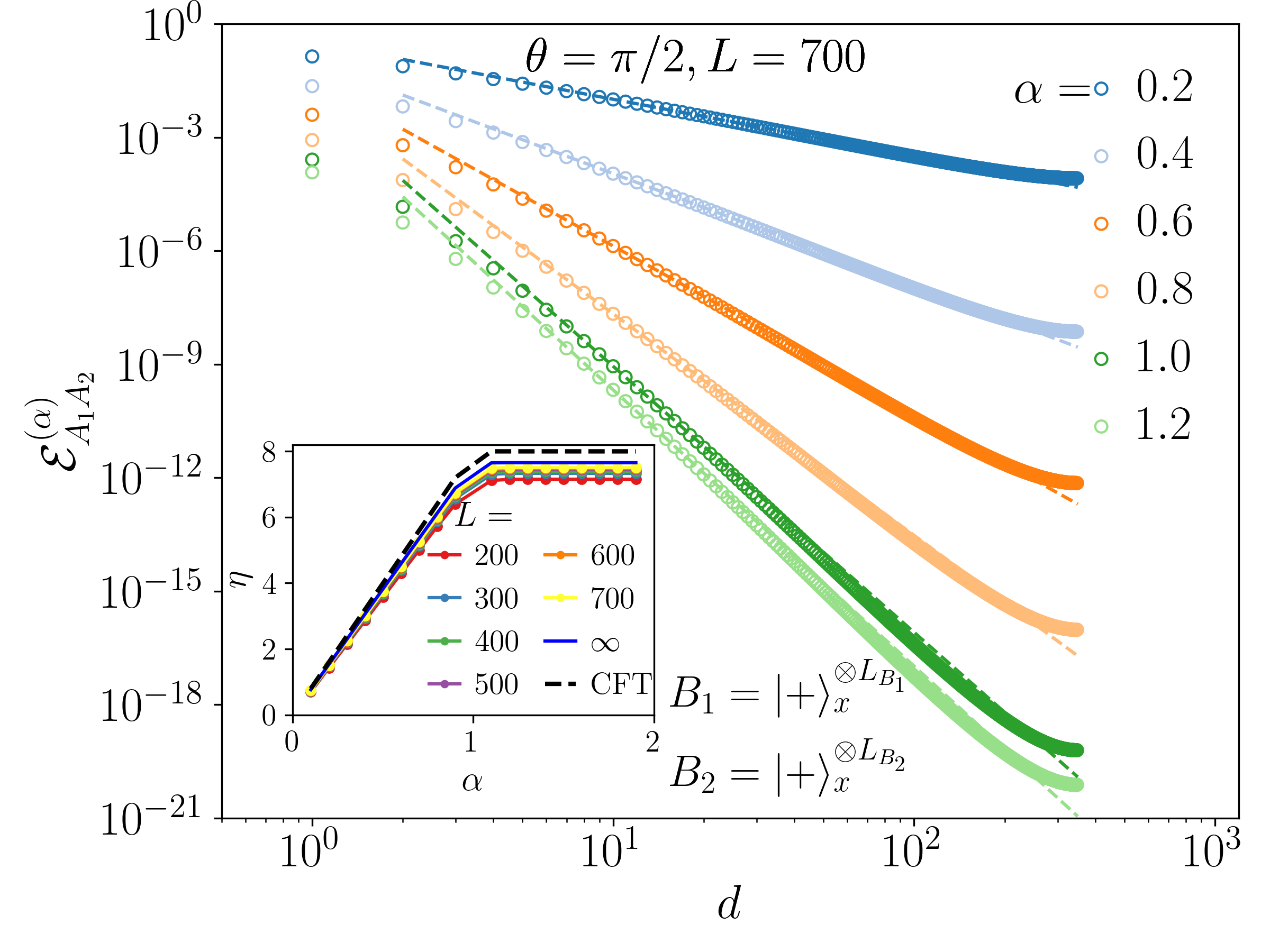}\put(-250,160){\textbf{(b)}}
\caption{\textbf{Transverse-field Ising chain at the critical point}. The main plots show the power-law decay of $\mathcal{E}_{A_1A_2}^{(\alpha)}$ as a function of distance $d$, while the insets display the corresponding decay exponent $\eta_\alpha$ for different values of $\alpha$. The dashed line represents the extrapolation to the thermodynamic limit ($L \to \infty$). The measurements are in the (a) $\sigma_z$  (b)  $\sigma_x$ basis.}
\label{fig:S_alpha_decay}
\end{figure}
Once the measurement is performed, the overall state collapses to one conditioned on the outcome \(\mathcal{S}_B\). The post measurement state of subsystem \(A\) is then given by 
\begin{equation}\label{psiA}
\ket{\psi_A^{(\mathcal{S}_B)}}
=\frac{1}{\sqrt{P_{\mathcal{S}_B}}}\sum_{\mathcal{S}_A} a_{(\mathcal{S}_A,\mathcal{S}_B)}(\mathbf{R},\boldsymbol{\phi},\boldsymbol{\theta},\boldsymbol{\alpha})\,\ket{\mathcal{S}_A}\,.
\end{equation}
After the measurement, the wave function collapses into a product state between \(A\) and \(B\), eliminating entanglement between the two subsystems. However, our interest lies in the residual (or induced) entanglement within subsystem \(A\) itself. To capture this, we further partition \(A\) into two disjoint parts, \(A_1\) and \(A_2\). The reduced density matrix of \(A_1\) is obtained by tracing over the degrees of freedom in \(A_2\):
\begin{equation}\label{rho_A1}
\rho_{A_1}^{(\mathcal{S}_B)} = \tr_{A_2}\left[ \ket{\psi_A^{(\mathcal{S}_B)}}\bra{\psi_A^{(\mathcal{S}_B)}} \right]\,.
\end{equation}

Finally, the entanglement entropy between \(A_1\) and \(A_2\) for a given measurement outcome \(\mathcal{S}_B\) can be quantified using the Rényi entanglement entropy, defined as
\begin{equation}
\mathcal{E}_{A_1}^{(\alpha, \mathcal{S}_B)} = \frac{1}{1 - \alpha} \ln \tr \left[ \left( \rho_{A_1}^{(\mathcal{S}_B)} \right)^\alpha \right]\,,
\end{equation}
where \(\alpha > 0\) and \(\alpha \neq 1\). In the limit \(\alpha \to 1\), the Rényi entropy reduces to the von Neumann entropy, given by
\begin{equation}\label{entanglement_entropy}
\mathcal{E}_{A_1}^{(\mathcal{S}_B)} = -\tr\left[\rho_{A_1}^{(\mathcal{S}_B)} \ln \rho_{A_1}^{(\mathcal{S}_B)}\right]\,.
\end{equation}
%
%
To place our findings in a broader context, we first present two key results derived from conformal field theory (CFT) applied to critical spin chains with short-range interactions. These results describe the scaling of post measurement entanglement entropy and the behavior of the decay exponent \(\eta(\alpha)\) as a function of the Rényi index \(\alpha\)~\cite{R2016}:

\begin{equation}\label{eq:S_entanglement_decay_general}
\mathcal{E}_{A_1}^{(\alpha, \mathcal{S}_B)}(d) \sim d^{-\eta(\alpha)},
\end{equation}
where $d$ denotes the distance between the two disjoint subsystems $A_1$ and $A_2$, and the sizes of these subsystems satisfy \(|A_1|, |A_2| \ll |B|\). The asymptotic exponent is given by
\begin{equation}\label{eq:eta_alpha_general}
\eta(\alpha) =
\begin{cases}
4\alpha \Delta_1, & \text{if } \alpha < 1 \\
4\Delta_1, & \text{if } \alpha \geq 1
\end{cases}
\end{equation}
Here, \(\Delta_1\) is the scaling dimension associated with the operator that creates the post-measurement state. Its value is determined by the boundary condition imposed by the configuration $\mathcal{S}_B$ from a renormalization group perspective and must be computed explicitly on the discrete spin chain.
These expressions hold for general spin chains at criticality, provided the system admits a CFT description.
Although the value of \(\Delta_1\) depends on the choice of measurement basis, the outcome of measurement  and the underlying critical model it still shows some level of universality that we will study here. In the critical transverse-field Ising chain, the scaling dimension \(\Delta_1\) has been  numerically studied in the \(\sigma^z\) basis~\cite{R2016}. Determining \(\Delta_1\) in an arbitrary measurement basis is, in general, a difficult task. However, our framework provides a practical way to extract this quantity numerically for any Pauli basis. We apply this method to several measurement bases; in particular, we present more detailed numerical results for the \(\sigma^x\) basis.

\medskip

 The Hamiltonian of the transverse-field Ising model is given by
\begin{equation}
H = -\sum_{i=1}^{L} \left( J\sigma_i^x \sigma_{i+1}^x + h\sigma_i^z \right),
\end{equation}
where \(L\) is the system size and the critical point occurs at $h=J$. The corresponding $\mathbf{R}$-matrix for the critical Ising chain is presented in \cite{TKR2024}. We now examine the distance-dependent decay of \(\mathcal{E}_{A_1}^{(\alpha, \mathcal{S}_B)}\), as defined in Eqs.~\eqref{rho_A1}--\eqref{entanglement_entropy}. 

Figure~\ref{fig:S_alpha_decay} displays the behavior of \(\mathcal{E}_{A_1}^{(\alpha, \mathcal{S}_B)}(d)\) for the measurement outcomes \(|B_1\rangle = |+\rangle^{\otimes L_{B_1}}\) and \(|B_2\rangle = |+\rangle^{\otimes L_{B_2}}\), in both the \(\sigma^z\) and \(\sigma^x\) bases. In both cases, the entanglement entropy exhibits a clear power-law decay with distance, in accordance with Eq.~\eqref{eq:S_entanglement_decay_general}. The extracted scaling dimensions are \(\Delta_1 = 1/2\) for the \(\sigma^z\) basis and \(\Delta_1 = 2\) for the \(\sigma^x\) basis.


A broader set of measurement outcomes, along with their corresponding decay exponents $\eta(\alpha)$, is summarized in Table~\ref{Table:1}. In what follows we always take $\phi\in[0,\pi/2)$; within this range all of the decay exponents we obtain are independent of $\phi$. For all configurations except configuration~4, any measurement basis with $\theta\in(0,\pi/2]$ yields the same decay exponent $\Delta_1$ as the special case $\theta=\pi/2$ (the $\sigma^x$ basis). By contrast, in configuration~4 one finds $\Delta_1 = 2$ for all $\theta\in(0,\pi/2)$. We suggest that similar exponents can be obtained for a broader range of crystal configurations. For instance, if the base of a crystal configuration has a definite magnetization, we anticipate it will behave similarly to the all-positive or all-negative configurations. Conversely, when the base has zero net magnetization, we expect its behavior to resemble that of the alternating $(+-)$ base configuration.

 These results underscore the versatility of our approach in characterizing post-measurement entanglement in critical spin chains. Although our primary focus has been on critical spin chains to facilitate comparison with conformal field theory predictions, our approach can readily be applied away from criticality. In such non-critical regimes, one observes an exponential decay of entanglement with increasing distance.

\begin{widetext}
\begin{center}
\begin{table}[h]
\centering
\renewcommand{\arraystretch}{1.9}
\rowcolors{1}{cyan!15!white}{white} 

\begin{tabular}{|c|c|c|c|}
    \hline
    \rowcolor{cyan!50!white} 
    \textbf{} & \textbf{Measurement Output $B_1$} & \textbf{Measurement Output $B_2$} & \textbf{Scaling dimension $\boldsymbol{(\Delta_1)}$} \\
    \hline
    1 & $|+\rangle_z^{\otimes L_{B_1}}$ & $|+\rangle_z^{\otimes L_{B_2}}$ & 1/2 \\
    2 & $|+\rangle_x^{\otimes L_{B_1}}$ & $|+\rangle_x^{\otimes L_{B_2}}$ & 2 \\
    3 & $|+\rangle_x^{\otimes L_{B_1}}$ & $|-\rangle_x^{\otimes L_{B_2}}$ & 1 \\
    4 & $|\!+\!-\rangle_x^{\otimes \frac{L_{B_1}}{2}}$ & $|\!+\!-\rangle_x^{\otimes \frac{L_{B_2}}{2}}$ & 1/2 \\
    5 & $|\!+\!-\rangle_x^{\otimes \frac{L_{B_1}}{2}}$ & $|+\rangle_x^{\otimes L_{B_2}}$ & 1 \\
    \hline
\end{tabular}
\caption{Scaling dimensions \(\Delta_1\) corresponding to different measurement outcome configurations on subsystems \(B_1\) and \(B_2\).}
\label{Table:1}
\end{table}
\end{center}
\end{widetext}
\section{Conclusion}

In this work, we derived an explicit Pfaffian formula for computing the amplitudes of fermionic Gaussian pure states in any Pauli basis, along with a recursive relation that connects amplitudes across different qubit sizes. Together, these results establish a robust computational framework that will be valuable for addressing complex quantum information problems, including formation probabilities, global entanglement, and quantum tomography. For illustration, we calculated the post-measurement entanglement entropy in transverse field Ising chain, uncovering scaling behavior consistent with conformal field theory. Our numerical results yield the precise decay exponents, demonstrating a degree of universality.

Although our analysis exclusively focuses on pure fermionic Gaussian states, it is important to note that many NISQ-era experiments operate under regimes where noise levels are sufficiently low—either naturally or via error mitigation—so that effective purity is maintained over the relevant timescales. In this context, fermionic Gaussian states serve not only as a theoretically tractable subclass of quantum states but also as practically relevant benchmarks for assessing quantum hardware performance~\cite{Gluza2018,Gorman2022,Mele2025}. Future work will consider extending these techniques to include mixed-state dynamics and explicit noise models.

Applying these formulas to specific quantum systems is a natural next step, expected to advance both theoretical and experimental research in Gaussian state analysis. Open questions remain, especially regarding extensions to Gaussian mixed states, where finding similar explicit formulas would broaden the applicability of Gaussian states in many-body physics and quantum computation. The methods introduced here offer essential tools for enhanced state analysis and practical applications in these fields.

{\bf Acknowledgements.}
We thank D. Brod for  discussion. MAR thanks CNPq and FAPERJ (Grant
No. E-26/210.062/2023) for partial support. The authors gratefully acknowledge the resources on the LiCCA HPC cluster of the University of Augsburg. The LiCCA cluster is co-funded by the Deutsche Forschungsgemeinschaft (DFG, German Research Foundation) – Project-ID 499211671.

\newpage

\appendix

\onecolumngrid
\section{Base configuration change in Generic Gaussian pure states}\label{sec:baseconfigurationchange}

The Gaussian pure state defined as \eqref{GPS-general-pfaffinho} which has the following remarkable property which is first presented in \cite{TKR2024}: consider the state $\ket{\bold{R},\mathcal{C}}$, then as far as the configuration $\ket{\mathcal{C}'}$ has non-zero amplitude one can always find an $\bold{R}'$ matrix such that:

\begin{equation}{\label{R matrix changes}}
\ket{\bold{R},\mathcal{C}} =\ket{\bold{R}',\mathcal{C}'}.  
\end{equation}
To get the matrix $\bold{R}'$ from $\bold{R}$ and the configurations ${\ket{\mathcal{C}}}={\ket{n_1,n_2,...,n_L}}$ and ${\ket{\mathcal{C}'}}={\ket{n'_1,n'_2,...,n'_L}}$ one can do the following: First consider the set of configurations ${\ket{\mathcal{I}'}}={\ket{m'_1,m'_2,...,m'_L}}$ that can be obtained from ${\ket{\mathcal{C}'}}$ by flipping just two spins. Then the elements of the matrix $\bold{R}'$, i.e. $r'_{ij}$, can be obtained as follows:

\begin{equation}{\label{R' matrix elements}}
  r'_{ij}=\text{sgn}(\mathcal{C},\mathcal{C}')\frac{\text{sgn} (\mathcal{C},\mathcal{I}')}{\text{sgn}(\mathcal{C}',\mathcal{I}')} \frac{\bold{pf}\ \bold{R}_{\mathcal{CI'}}}{\bold{pf}\ \bold{R}_{\mathcal{CC'}}},
\end{equation}
where the sets

\begin{eqnarray}{\label{the sets}}
  \mathcal{CI'}&=&\{\forall j|n_j-m'_j\neq0\},\\
  \mathcal{CC'}&=&\{\forall j|n_j-n'_j\neq0\},
\end{eqnarray}
and the signs $\text{sgn}(\mathcal{C},\mathcal{I}')$ and $\text{sgn}(\mathcal{C},\mathcal{C}')$ are defined as:

\begin{eqnarray}\label{GPS-general-sign} 
\text{sgn}(\mathcal{C}, \mathcal{I}') &=& \prod_{i=2}^L (-1)^{|n_i - m'_i| \sum_{j < i} n_j},\\
\text{sgn}(\mathcal{C}, \mathcal{C}') &=& \prod_{i=2}^L (-1)^{|n_i - n'_i| \sum_{j < i} n_j}.
\end{eqnarray}
For instance, when $\ket{\mathcal{C}} = \ket{\mathbf{0}}$, the signs in \eqref{R' matrix elements} and \eqref{GPS-general-pfaffinho} will vanish, and equation \eqref{GPS-general-pfaffinho} simpify to equation \eqref{GPS0-pfaffinho}.

\section{Proof of Theorem 1}\label{sec:theorem1}

In this Appendix, we outline the derivation of the main equation of Theorem \ref{theorem1}. The rules to determine the contributions of the angles \( \alpha \) and \( \phi \) are straightforward to identify. The angle \( \alpha \) appears as a phase factor and depends on the number of down spins. The angle \( \phi \) consistently appears as a phase factor multiplied with the elements of the matrix \( \boldsymbol{R} \) justifying the definition of $\bold{R}^{\boldsymbol{\phi}}$ with elements $r^{\boldsymbol{\phi}}_{nm}=e^{i(\phi_n+\phi_m)}r_{nm}$. The main challenge lies in understanding how the angle \( \theta \) contributes to the amplitudes.
The key idea is based on the observation that when expanding \( \boldsymbol{U}_{(\boldsymbol{\phi}, \boldsymbol{\theta}, \boldsymbol{\alpha})} \ket{\psi}_z \), each element of this vector consists of an exponential number of terms, i.e. pfaffinhos. One can first try to determine the coefficient of \(\bold{pf} \bold{R}^{\phi}_{\mathcal{I}}\) in each amplitude with respect to the trigonometric functions. Note that in the $\sigma^z$ basis, this is the amplitude of the configurations with all the spins up in the set $\mathcal{I}$, i.e. $\ket{\mathcal{S}_{\mathcal{I}}^z}$. Direct inspection shows that the coefficient is the following:
 \begin{eqnarray}\label{trigonometric}
  \prod_{j\in \mathcal{S}_{\mathcal{I}}}\cos\frac{\theta_j}{2}\prod_{j\in \overline{\mathcal{S}_{\mathcal{I}}}}\sin\frac{\theta_j}{2},
\end{eqnarray}
where $\mathcal{S}_{\mathcal{I}}$ is the set of sites that spins in  $\ket{\mathcal{S}}$ and  $\ket{\mathcal{S}_{\mathcal{I}}^z}$ are in the same direction. After factoring out $\prod\limits_{j\in \mathcal{S}^+}\cos\frac{\theta_j}{2}\prod\limits_{j\in \mathcal{S}^-}\sin\frac{\theta_j}{2}$ and rearranging the terms and summing them one can reach to the equation presented in the theorem \ref{theorem1}. The formula clearly reminds us of the Lieb's formula for pfaffians \cite{Lieb2016} which states the following: Given an antisymmetric matrix of even size $L$ and a collection of weights $\{\lambda_1,\lambda_2,...,\lambda_L\}$ define the matrix
\begin{eqnarray}\label{Rlambda}
\textbf{R}_{nm}(\lambda_1,...,\lambda_L)=R_{nm}-(-1)^{n+m}\lambda_i\lambda_j, \hspace{0.5cm}j>i.
\end{eqnarray}
Then we have:
\begin{eqnarray}\label{Lieb}
\bold{pf}\textbf{R}(\lambda_1,...,\lambda_L)=\sum_{k=0}^{\frac{L}{2}}\sum_{|\bar{I}|=L-2k}\bold{pf}\bold{R}_{\bar{I}}\prod_{j\in I}\lambda_j
\end{eqnarray}
It is possible to see that the amplitude can be found by applying the Lieb's formula with  $\lambda_j=\tan\frac{\theta_j}{2}$($\lambda_j=-\cot\frac{\theta_j}{2}$) when the spins are up (down).

The following theorem generalizes Lieb's formula to matrices of odd dimensions, enabling the calculation of amplitudes for these cases as well.

\begin{theorem}[L odd] When the matrix $\bold{R}$ has an odd dimension we add a zero row and column to the antisymmetric matrix $\bold{R}$, along with a set of weights $\{\lambda_1,\lambda_2,...,\lambda_{L'}\}$, 
and define the matrix $\textbf{R}^{ex}(\lambda_1,...,\lambda_{L'})$ as before, 
then we have: 
\begin{eqnarray}\label{Lieb odd case}
\bold{pf}\textbf{R}^{ex}(\lambda_1,...,\lambda_{L'})=\sum_{k=0}^{\frac{L'}{2}}\sum_{|\bar{I}|=L'-2k}\bold{pf}\bold{R}^{ex}_{\bar{I}}\prod_{j\in I}\lambda_j, \hspace{1cm} |{I}|+|\bar{I}|=L',
\end{eqnarray}
where $L'=L+1$, $\lambda_{L+1}=1$, and
\begin{eqnarray}
\textbf{R}^{ex}_{nm}(\lambda_1,...,\lambda_{L'})=R^{ex}_{nm}-(-1)^{n+m}\lambda_i\lambda_j, \hspace{1cm}j>i.
\end{eqnarray}
The formula \eqref{Lieb odd case} can also be expressed in terms of the $\bold{R}$ matrix as follows:
\begin{eqnarray}
\bold{pf}\textbf{R}^{ex}(\lambda_1,...,\lambda_{L'})=\sum_{k=0}^{L}\sum_{|\bar{I}|=L-k}\bold{pf}\bold{R}_{\bar{I}}\prod_{j\in I}\lambda_j, \hspace{1cm} |{I}|+|\bar{I}|=L.
\end{eqnarray}

\end{theorem}
The theorem can be proven either by following Lieb's original approach or by utilizing the Berezin integral representation over Grassmann variables for the pffafian of the matrix $\textbf{R}(\lambda_1,...,\lambda_{L'})$, followed by expanding the terms dependent on the \(\lambda\) variables and performing the Berezin integrals.                   

This formula, with the appropriate inclusion of an ancillary qubit, can be extended to derive the amplitudes for an odd number of qubits in arbitrary Pauli bases. The ancillary qubit is constrained to the $\sigma^x$ basis and aligned in the same direction as the first qubit.

\section{Alternative  formulas for the amplitudes }\label{S3}

In this Appendix, we present alternative forms of the amplitude formula. First, note that the main formula of theorem \ref{theorem1} can be also written as follows:
\begin{eqnarray}\label{theorem 1, second version}
 a_{\mathcal{S}}(\boldsymbol{\phi},\boldsymbol{\theta},\boldsymbol{\alpha})=\frac{(-1)^{L(1-\bar{s}_1)/2}\sqrt{2}^{L\text{mod} \ 2} e^{-i\sum\limits_{j\in\mathcal{S}^-}\alpha_j}}{\mathcal{N}_{R}} \bold{pf} \bold{M}(\boldsymbol{\phi},\boldsymbol{\theta},\boldsymbol{\alpha}),
\end{eqnarray}
where the antisymmetric matrix $\bold{M}(\boldsymbol{\phi},\boldsymbol{\theta},\boldsymbol{\alpha})$ is defined for $m>n$ as:
\begin{eqnarray}\label{M,bsnsm}
\bold{M}_{nm}(\boldsymbol{\phi},\boldsymbol{\theta},\boldsymbol{\alpha})=
    b_{s_n,s_m}(\boldsymbol{\phi},\boldsymbol{\theta},\boldsymbol{\alpha}),
\end{eqnarray}
in which
\begin{eqnarray}
 b_{s_n,s_m}(\boldsymbol{\phi},\boldsymbol{\theta},\boldsymbol{\alpha})= \left(\cos\frac{\theta_n}{2}\right)^{\frac{1+\bar{s}_n}{2}}  \left(\cos\frac{\theta_m}{2}\right)^{\frac{1+\bar{s}_m}{2}} \left(\sin\frac{\theta_n}{2}\right)^{\frac{1-\bar{s}_n}{2}} 
 \left(\sin\frac{\theta_m}{2}\right)^{\frac{1-\bar{s}_m}{2}}r_{nm}e^{i(\phi_n+\phi_m)} \hspace{1cm}\nonumber\\ +(-1)^{n+m}(-1)^{\frac{\bar{s}_n+\bar{s}_m}{2}}\left(\sin\frac{\theta_n}{2}\right)^{\frac{1+\bar{s}_n}{2}} \left(\sin\frac{\theta_m}{2}\right)^{\frac{1+\bar{s}_m}{2}} \left(\cos\frac{\theta_n}{2}\right)^{\frac{1-\bar{s}_n}{2}} \left(\cos\frac{\theta_m}{2}\right)^{\frac{1-\bar{s}_m}{2}}.
\end{eqnarray}
For odd sizes $\bar{s}_{L+1}=\bar{s}_1$ and $\theta_{L+1}=\frac{\pi}{2}$ and $\alpha_{L+1}=0$.
The formula in \eqref{theorem 1, second version} can be simplified for the special cases when all spins are up or all spins are down, as follows:
\begin{eqnarray}
 a_{+...+}(\boldsymbol{\phi},\boldsymbol{\theta},\boldsymbol{\alpha})=\frac{\sqrt{2}^{L\text{mod} \ 2}}{\mathcal{N}_{R}} \bold{pf} \bold{M}(\boldsymbol{\phi},\boldsymbol{\theta},\boldsymbol{\alpha}),
\end{eqnarray}
where
\begin{eqnarray}
 b_{s_n,s_m}(\boldsymbol{\phi},\boldsymbol{\theta},\boldsymbol{\alpha})=\cos\frac{\theta_n}{2}  \cos\frac{\theta_m}{2} r_{nm}e^{i(\phi_n+\phi_m)} -(-1)^{n+m}\sin\frac{\theta_n}{2} \sin\frac{\theta_m}{2},
\end{eqnarray}
and
\begin{eqnarray}
 a_{-...-}(\boldsymbol{\phi},\boldsymbol{\theta},\boldsymbol{\alpha})=\frac{(-1)^L\sqrt{2}^{L\text{mod} \ 2} e^{-i\sum\limits_{j=1}^L\alpha_j}}{\mathcal{N}_{R}} \bold{pf} \bold{M}(\boldsymbol{\phi},\boldsymbol{\theta},\boldsymbol{\alpha}),
\end{eqnarray}
where
\begin{eqnarray}
 b_{s_n,s_m}(\boldsymbol{\phi},\boldsymbol{\theta},\boldsymbol{\alpha})= \sin\frac{\theta_n}{2}  \sin\frac{\theta_m}{2} r_{nm}e^{i(\phi_n+\phi_m)} -(-1)^{n+m}\cos\frac{\theta_n}{2} \cos\frac{\theta_m}{2}.
\end{eqnarray}
In the Table \ref{bsn,sm}, we also summarize the expressions for $b_{s_n,s_m}(\boldsymbol{\phi},\boldsymbol{\theta},\boldsymbol{\alpha})$ in different basis.
\begin{table}[h]
\centering
\renewcommand{\arraystretch}{1.9}
\rowcolors{1}{cyan!15!white}{white} 

\begin{tabular}{|c|c|}
    \hline
    \rowcolor{cyan!50!white} 
    \textbf{} & \textbf{Expression} \\
    \hline
    {$b_{s_n,s_m}^{zz}$}  & $(\frac{1+\bar{s}_n}{2})(\frac{1+\bar{s}_m}{2})r_{nm}-(-1)^{n+m}(\frac{1-\bar{s}_n}{2})(\frac{1-\bar{s}_m}{2})$ \\
    \hline
    $b_{s_n,s_m}^{xx}$  & $\frac{1}{2}\left(r_{nm}-(-1)^{n+m}\bar{s}_n\bar{s}_m\right)$ \\
    \hline
    $b_{s_n,s_m}^{yy}$  & $-\frac{1}{2}\left(r_{nm}+(-1)^{n+m}\bar{s}_n\bar{s}_m\right)$ \\
    \hline
    $b_{s_n,s_m}^{zx}$  & $\frac{1}{\sqrt{2}}\left[(\frac{1+\bar{s}_n}{2})r_{nm}+(-1)^{n+m}\bar{s}_m(\frac{1-\bar{s}_n}{2})\right]$ \\
    \hline
    $b_{s_n,s_m}^{zy}$  & $\frac{1}{\sqrt{2}}\left[(\frac{1+\bar{s}_n}{2})ir_{nm}+(-1)^{n+m}\bar{s}_m(\frac{1-\bar{s}_n}{2})\right]$ \\
    \hline
    $b_{s_n,s_m}^{xy}$  & $\frac{1}{2}\left(ir_{nm}-(-1)^{n+m}\bar{s}_n\bar{s}_m\right)$ \\
    \hline
\end{tabular}
\caption{$b_{s_n,s_m}(\boldsymbol{\phi},\boldsymbol{\theta},\boldsymbol{\alpha})$ in different basis.}\label{bsn,sm}
\end{table} 

To make the connection with the formulas that we presented in the computational basis in the main text which were based on pfaffinhos we first represent the antisymmetric matrix $\bold{M}$ in the $\sigma^z$ basis as follows:

\begin{eqnarray}
\bold{M}_{nm}=\begin{cases}
    r_{nm} \hspace{3cm} \bar{s}_n,\bar{s}_m=1,\\
    (-1)^{n+m+1} \hspace{1.85cm} \bar{s}_n,\bar{s}_m=-1,\\
    0 \hspace{3.4cm} \bar{s}_n=1,\bar{s}_m=-1 \ or \ \bar{s}_n=-1,\bar{s}_m=1.
\end{cases}
\end{eqnarray}
By expanding along rows other than $r_{nm}$, we can write:
\begin{eqnarray}
\bold{pf}[\bold{M}_{{n}{m}}]=\bold{pf}\bold{R}_{\mathcal{I}},
\end{eqnarray}
where we have used the property:
\begin{equation}
\textbf{pf}[A] = \sum_{j=1, j \neq i}^L (-1)^{i+j+1 + H(i-j)} a_{ij} \, \textbf{pf}[A_{\overline{ij}}],
\end{equation}
in which $a_{ij}$ is the elements of the matrix $A$, $\overline{ij}$ is the complement of $ij$, and $A_{\overline{ij}}$ is the matrix $A$ in which we removed the rows and columns $i$ and $j$, and $H(i-j)$ representing the Heaviside step function.\\

There is another version of the formula \eqref{M,bsnsm} that relates the amplitudes of \( L \)-qubit systems to those of two-qubit cases with adjusted angles $\boldsymbol{\theta}$. It has the following form:
\begin{eqnarray}
\bold{M}_{nm}(\boldsymbol{\phi},\boldsymbol{\theta},\boldsymbol{\alpha})=\begin{cases}
    \mathcal{N}_{R_{nm}}a_{s_n,s_m}(\boldsymbol{\phi},\boldsymbol{\theta},\boldsymbol{\alpha}) \hspace{4cm} n+m \ \text{odd},\\
    (-1)^{\frac{1+\bar{{s}}_n}{2}}\mathcal{N}_{R_{nm}}a_{s_n,s_m}(\boldsymbol{\phi},2\pi-\theta_n,\theta_m,\boldsymbol{\alpha}) \hspace{1cm} n+m \ \text{even},
\end{cases}
\end{eqnarray}
in which 
\begin{eqnarray}
b_{s_n,s_m}(\boldsymbol{\phi},\boldsymbol{\theta},\boldsymbol{\alpha})={\mathcal{N}_{R_{nm}}} a_{s_n,s_m}(\boldsymbol{\phi},\boldsymbol{\theta},\boldsymbol{\alpha}), \hspace{2cm} \mathcal{N}_{R_{nm}}=\det\left(\bold{I}+\bold{R}^{\dagger}_{nm}.\bold{R}_{nm}\right)^{\frac{1}{4}}=(1+|r_{nm}|^2)^\frac{1}{2},
\end{eqnarray}
where $\bold{R}_{nm}$ is the matrix $\bold{R}$ in which we keep the rows and columns $n,m$. One can also write the formula \eqref{M,bsnsm} in another way by just adjusting the angles $\boldsymbol{\phi}$ as follows:
\begin{eqnarray}
\bold{M}_{nm}(\boldsymbol{\phi},\boldsymbol{\theta},\boldsymbol{\alpha})=\begin{cases}
    {\mathcal{N}_{R_{nm}}}a_{s_n,s_m}(\boldsymbol{\phi},\boldsymbol{\theta},\boldsymbol{\alpha}) \hspace{4cm} n+m \ \text{odd},\\
    -{\mathcal{N}_{R_{nm}}}a_{s_n,s_m}(\boldsymbol{\phi}+\frac{\pi}{2},\boldsymbol{\theta},\boldsymbol{\alpha}) \hspace{3cm} n+m \ \text{even}.
\end{cases}
\end{eqnarray}

\section{Relations among amplitudes}\label{Relationsamongamplitudes}
In this Appendix, we present a formula that expresses the amplitudes of a system in terms of the amplitudes corresponding to configurations where only two spins are aligned in the positive direction. Since the Gaussian pure states are characterized by the $\frac{L(L-1)}{2}$ elements of the matrix $\mathbf{R}$, one would expect to have the same number of independent amplitudes. For example, when $\theta \neq \frac{\pi}{2}$, all amplitudes can be expressed in terms of the amplitudes corresponding to configurations where only two spins are in the positive direction. As an illustrative example, we consider the amplitude for a system of size $L=4$ in the $\sigma^{z}$ basis, in which the angles $\boldsymbol{\theta}$ and $\boldsymbol{\phi}$ are zero. Then we have:
\begin{eqnarray}\label{Relations among amplitudes in sigma z basis}
a_{\ket{\downarrow \downarrow \downarrow \downarrow}} a_{\ket{\uparrow \uparrow \uparrow \uparrow}}=a_{\ket{\uparrow \uparrow \downarrow \downarrow}} a_{\ket{\downarrow \downarrow \uparrow \uparrow}} -a_{\ket{\uparrow \downarrow \uparrow \downarrow}} a_{\ket{\downarrow \uparrow \downarrow \uparrow}}+a_{\ket{\uparrow \downarrow \downarrow \uparrow}} a_{\ket{\downarrow \uparrow \uparrow \downarrow}}.
\end{eqnarray}
We showed that the Gaussian pure state in the computational basis ($\sigma^z$ basis) can be written as:
\begin{equation}
    \ket{\bold{R},\mathbf{0}} = \frac{1}{\mathcal{N}_R}\sum_{\mathcal{I}}\bold{pf}\ \bold{R}_{\mathcal{I}}{\ket{\mathcal{I}}},
\end{equation}
which coincides with the results of \eqref{Relations among amplitudes in sigma z basis}. Note that we consider the same basis for all qubits. The second example is $(\phi,\frac{\pi}{2},0)$ basis, which describes the $\ket{\bold{R},\bold{0}} $  in the domain wall basis of the $(\phi,\frac{\pi}{2},0)$ basis. To understand the domain wall basis, consider an arbitrary spin sequence such as \(\ket{+-+--\dots-+}\) in the \((\phi, \frac{\pi}{2}, 0)\) basis. The corresponding domain wall representation of this sequence is \(\ket{\tilde{1}\tilde{1}\tilde{1}\tilde{0}\dots\tilde{1}\tilde{0}}\). Note that the domain wall structure also accounts for the boundary condition, specifically the domain wall between the last spin and the first spin in the sequence. Importantly, reversing the direction of all spins in the \((\phi, \frac{\pi}{2}, 0)\) basis results in the same domain wall configuration. This symmetry implies that the mapping from spin configurations to domain wall configurations is two-to-one. In \cite{TKR2024} it is shown that the state in the $(\phi,\frac{\pi}{2},0)$ basis will be:
\begin{equation}\label{phi,pi2,0}
    \ket{\bold{R},\bold{0}} = \frac{1}{\sqrt{2}\mathcal{N}_{\tilde{R}^{\phi}}}\sum_{\mathcal{S}}sgn(\mathcal{S})\bold{pf}\ (\tilde{\bold{R}}^{\phi})_{\mathcal{S}}{\ket{\mathcal{S}}}_{\phi},
\end{equation}
where the $
\mathbf{\tilde{R}}^{\phi}$ can be found as follows:
\begin{equation}{\label{R bar phi}}
    \tilde{\bold{R}}^{\phi} = (\bold{I}+\bold{W}^{\phi}.\bold{P})(\bold{W}^{\phi}.\bold{P}-\bold{I})^{-1},
\end{equation}
in which
\begin{eqnarray}{\label{H phi}}
\bold{W}^{\phi} = (\bold{R}^{\phi}-\bold{I}).(\bold{R}^{\phi}+\bold{I})^{-1}, \ \ \ \ \ \ \ \ \ \ \bold{R}^{\phi}=e^{2i\phi}\bold{R},
\end{eqnarray}
and
\begin{equation}\label{eq: P matrix}
\begin{split}
\bold{P}= \left(
\begin{array}{ccccccc}
 0 & 0 & 0 & 0 & \hdots & 0 & 1 \\
 -1 & 0 & 0 & 0 & \hdots & 0 & 0 \\
0 & -1 & 0 & 0 & \hdots & 0 & 0 \\
\vdots& \vdots& \vdots& \vdots& \ddots& \vdots& \vdots \\
 0 & 0 & 0 & 0 & \hdots & 0 & 0 \\
0 & 0 & 0 & 0 & \hdots & 0 & 0 \\
0 & 0 & 0 & 0 & \hdots & -1 & 0 \\
\end{array}
\right).
\end{split}
\end{equation}
In the formula \eqref{phi,pi2,0}, ${\ket{\mathcal{S}}}_{\phi}$ as before is a sequence of $+$ and $-$ in the $\phi$ basis and the $(\tilde{\bold{R}}^{\phi})_{\mathcal{S}}$ is a submatrix of the matrix $\tilde{\bold{R}}^{\phi}$ in which we first find the domain wall configuration of $\mathcal{S}$ and then we keep the rows and columns corresponding to the sites that there is a domain wall. The $sgn(\mathcal{S})$ is simply +1(-1) for even(odd) number of $-$ in the sequence. As an example, we consider the amplitudes for a system of size $L=4$ in the $(\phi,\frac{\pi}{2},0)$ basis, then we can write:
\begin{eqnarray}
a_{\ket{++++}} a_{\ket{+-+-}}=a_{\ket{+-++}} a_{\ket{+++-}} -a_{\ket{+--+}} a_{\ket{++--}}+a_{\ket{+---}} a_{\ket{++-+}},\\
a_{\ket{----}} a_{\ket{-+-+}}=a_{\ket{-+--}} a_{\ket{---+}} -a_{\ket{-++-}} a_{\ket{--++}}+a_{\ket{-+++}} a_{\ket{--+-}},
\end{eqnarray}
which is equivalent to the results of the \eqref{phi,pi2,0}. For the general case, where $\theta \neq \frac{\pi}{2}$, a relationship between the amplitudes still holds; however, its derivation is significantly more complex and challenging to explicitly demonstrate.

\section{Alternative  formula for the recursive formula}\label{Alternativeformulafortherecursiveformula}
In this Appendix, we first prove Theorem \ref{theorem2recursiveformula}, which establishes that the amplitudes of $L$-qubit systems can be expressed in terms of the amplitudes of 2-qubit and $L-2$-qubit systems. Subsequently, we present an alternative formula for the recursive formula. Theorem \ref{theorem2recursiveformula} can be derived directly from Theorem \ref{theorem1} by employing the fundamental properties of the Pfaffian. This derivation involves decomposing the Pfaffian in Theorem \ref{theorem1} into a sum over pairwise contributions and recursive terms, where the angles $\boldsymbol{\theta}$ and $\boldsymbol{\phi}$ are adjusted to accommodate the reduced subsystems. Thus the $b_{\mathcal{S}}(\boldsymbol{\phi},\boldsymbol{\theta},\boldsymbol{\alpha})$ can be written as:
\begin{eqnarray}
\hspace{10cm}\nonumber\\\sum_{j=1}^{L'/2} b_{s_1,s_{2j}}(\boldsymbol{\phi},\boldsymbol{\theta},\boldsymbol{\alpha})b_{\mathcal{S}/\{s_1,s_{2j}\}}(\boldsymbol{\phi},\bar{\boldsymbol{\theta}}^j,\boldsymbol{\alpha})
 -\sum_{j=1}^{L'/2-1} b_{s_1,s_{2j+1}}(\boldsymbol{\phi}+\frac{\pi}{2},\boldsymbol{\theta},\boldsymbol{\alpha})b_{\mathcal{S}/\{s_1,s_{2j+1}\}}(\boldsymbol{\phi},2\pi-\bar{\boldsymbol{\theta}}^j,\boldsymbol{\alpha}).
\end{eqnarray}
This formula connects the amplitude of \( L \)-qubit systems to those of \( 2 \)-qubit and \( L-2 \)-qubit systems. However, adjusting the angles introduces additional sign factors, which are accounted for by assigning a spin parity sign to each term in the recursive sum, as shown below:
\begin{eqnarray}
 b_{\mathcal{S}}(\boldsymbol{\phi},\boldsymbol{\theta},\boldsymbol{\alpha})=\hspace{10cm}\nonumber\\\sum_{j=1}^{L'/2} P(\bar{A}_{2j})b_{s_1,s_{2j}}(\boldsymbol{\phi},\boldsymbol{\theta},\boldsymbol{\alpha})b_{\mathcal{S}/\{s_1,s_{2j}\}}(\boldsymbol{\phi},\bar{\boldsymbol{\theta}}^j,\boldsymbol{\alpha})
 -\sum_{j=1}^{L'/2-1} P(A_{2j})b_{s_1,s_{2j+1}}(\boldsymbol{\phi}+\frac{\pi}{2},\boldsymbol{\theta},\boldsymbol{\alpha})b_{\mathcal{S}/\{s_1,s_{2j+1}\}}(\boldsymbol{\phi},2\pi-\bar{\boldsymbol{\theta}}^j,\boldsymbol{\alpha}).
\end{eqnarray}
Following this, we summarize the expressions for $b_{\mathcal{S}}(\boldsymbol{\phi},\boldsymbol{\theta},\boldsymbol{\alpha})$ in Theorem \ref{theorem2recursiveformula} for different system sizes in Table \ref{bS}. Additionally, we provide a new recursive formula that connects the amplitudes of \( L \)-qubit systems to \( L-2 \)-qubit systems, similar to Theorem \ref{theorem2recursiveformula} but featuring different angle relations. One such alternative form is as follows:
\begin{eqnarray}
 b_{\mathcal{S}}(\boldsymbol{\phi},\boldsymbol{\theta},\boldsymbol{\alpha})=\sum_{j=1}^{L'/2} P(\bar{A}_{2j})b_{s_1,s_{2j}}(\boldsymbol{\phi},\boldsymbol{\theta},\boldsymbol{\alpha})b_{\mathcal{S}/\{s_1,s_{2j}\}}(\boldsymbol{\phi},\bar{\boldsymbol{\theta}}^j,\boldsymbol{\alpha})\hspace{3cm}\nonumber\\
 +\sum_{j=1}^{L'/2-1} (-1)^{\frac{1+\bar{s}_1}{2}}P(A_{2j})b_{s_1,s_{2j+1}}(\boldsymbol{\phi},2\pi-\theta_1,\theta_{2j+1},\boldsymbol{\alpha})b_{\mathcal{S}/\{s_1,s_{2j+1}\}}(\boldsymbol{\phi},2\pi-\bar{\boldsymbol{\theta}}^j,\boldsymbol{\alpha}),
\end{eqnarray}
where the notation is the same as the theorem \ref{theorem2recursiveformula}. $b_{s_n,s_{m}}=\mathcal{N}_{R_{nm}}a_{s_n,s_{m}}$ and $b_{\mathcal{S}/\{s_n,s_{m}\}}=\mathcal{N}_{R_{\overline{nm}}}a_{\mathcal{S}/\{s_n,s_{m}\}}$, in which 
$\overline{nm}$ is the complement of the $nm$. The normalization $\mathcal{N}_{R_{nm}}=\det\left(\bold{I}+\bold{R}^{\dagger}_{nm}.\bold{R}_{nm}\right)^{\frac{1}{4}}$ and $\mathcal{N}_{{R}_{\overline{nm}}}=\det\left(\bold{I}+\bold{R}_{\overline{nm}}^{\dagger}.\bold{R}_{\overline{nm}}\right)^{\frac{1}{4}}$, where $\bold{R}_{nm}$ is the matrix $\bold{R}$ in which we keep the rows and columns $n,m$ and $\bold{R}_{\overline{nm}}$ is the matrix $\bold{R}$ in which we remove the rows and columns $n,m$. $P(D)=\prod\limits_{k\in D}\bar{s}_k$ as the parity of spins in the region $D$, and $A_{2j}=\{2,3,...,2j\}$, $\bar{A}_{2j}=\{2j+1,2j+2,...,L'\}$ and the angle can be found as:
\begin{eqnarray}
\bar{\boldsymbol{\theta}}^j=\begin{cases}
    2\pi-\theta_k\hspace{0.5cm}k\in\bar{A}_{2j},\\
    \theta_k\hspace{1.3cm}k\in A_{2j}.
\end{cases}
\end{eqnarray}
For odd sizes $\bar{s}_{L+1}=\bar{s}_1$ and $\theta_{L+1}=\frac{\pi}{2}$ and $\alpha_{L+1}=0$. 
\begin{table}[h]
\centering
\renewcommand{\arraystretch}{1.9}
\rowcolors{1}{cyan!15!white}{white} 
\begin{tabular}{|c|c|}
    \hline
    \rowcolor{cyan!50!white} 
    \text{Qubit Number} & $b_{\mathcal{S}}(\boldsymbol{\phi},\boldsymbol{\theta},\boldsymbol{\alpha})$ \\
    \hline
    $3$  & $\bar{s}_3\bar{s}_1 b_{s_1,s_2}(\boldsymbol{\phi},\boldsymbol{\theta},\boldsymbol{\alpha}) b_{s_3,s_1}(\boldsymbol{\phi},2\pi-\theta_3,\frac{3\pi}{2},\boldsymbol{\alpha})$ \\ & $+b_{s_1,s_1}(\boldsymbol{\phi},\theta_1,\frac{\pi}{2},\boldsymbol{\alpha}) b_{s_2,s_3}(\boldsymbol{\phi},\boldsymbol{\theta},\boldsymbol{\alpha})-\bar{s}_2b_{s_1,s_3}({\phi}_1+\frac{\pi}{2},{\phi}_3+\frac{\pi}{2},\boldsymbol{\theta},\boldsymbol{\alpha}) b_{s_2,s_1}(\boldsymbol{\phi},2\pi-\theta_2,\frac{\pi}{2},\boldsymbol{\alpha})$ \\
    \hline
    $4$  & $\bar{s}_3\bar{s}_4 b_{s_1,s_2}(\boldsymbol{\phi},\boldsymbol{\theta},\boldsymbol{\alpha}) b_{s_3,s_4}(\boldsymbol{\phi},2\pi-\theta_3,2\pi-\theta_4,\boldsymbol{\alpha})$ \\ 
         & +$b_{s_1,s_4}(\boldsymbol{\phi},\boldsymbol{\theta},\boldsymbol{\alpha}) b_{s_2,s_3}(\boldsymbol{\phi},\boldsymbol{\theta},\boldsymbol{\alpha})-\bar{s}_2b_{s_1,s_3}({\phi}_1+\frac{\pi}{2},{\phi}_3+\frac{\pi}{2},\boldsymbol{\theta},\boldsymbol{\alpha}) b_{s_2,s_4}(\boldsymbol{\phi},2\pi-\theta_2,2\pi-\theta_4,\boldsymbol{\alpha})$ \\
    \hline
    {}  & $\bar{s}_3\bar{s}_4\bar{s}_5\bar{s}_1 b_{s_1,s_2}(\boldsymbol{\phi},\boldsymbol{\theta},\boldsymbol{\alpha}) b_{s_3,s_4,s_5,s_1}(\boldsymbol{\phi},2\pi-\theta_3,2\pi-\theta_4,2\pi-\theta_5,\frac{3\pi}{2},\boldsymbol{\alpha})$ \\ $5$
         & $+\bar{s}_5\bar{s}_1 b_{s_1,s_4}(\boldsymbol{\phi},\boldsymbol{\theta},\boldsymbol{\alpha}) b_{s_2,s_3,s_5,s_1}(\boldsymbol{\phi},\theta_2,\theta_3,2\pi-\theta_5,\frac{3\pi}{2},\boldsymbol{\alpha})+b_{s_1,s_1}(\boldsymbol{\phi},\theta_1,\frac{\pi}{2},\boldsymbol{\alpha}) b_{s_2,s_3,s_4,s_5}(\boldsymbol{\phi},\boldsymbol{\theta},\boldsymbol{\alpha})$ \\
         & $-\bar{s}_2b_{s_1,s_3}({\phi}_1+\frac{\pi}{2},{\phi}_3+\frac{\pi}{2},\boldsymbol{\theta},\boldsymbol{\alpha})b_{s_2,s_4,s_5,s_1}(\boldsymbol{\phi},2\pi-\theta_2,\theta_4,\theta_5,\frac{\pi}{2},\boldsymbol{\alpha})$\\ &$-\bar{s}_2\bar{s}_3\bar{s}_4b_{s_1,s_5}({\phi}_1+\frac{\pi}{2},{\phi}_5+\frac{\pi}{2},\boldsymbol{\theta},\boldsymbol{\alpha})b_{s_2,s_3,s_4,s_1}(\boldsymbol{\phi},2\pi-\theta_2,2\pi-\theta_3,2\pi-\theta_4,\frac{\pi}{2},\boldsymbol{\alpha})$ \\
    \hline
      & $\bar{s}_3\bar{s}_4\bar{s}_5\bar{s}_6 b_{s_1,s_2}(\boldsymbol{\phi},\boldsymbol{\theta},\boldsymbol{\alpha}) b_{s_3,s_4,s_5,s_6}(\boldsymbol{\phi},2\pi-\theta_3,2\pi-\theta_4,2\pi-\theta_5,2\pi-\theta_6,\boldsymbol{\alpha})$ \\ $6$
         & $+\bar{s}_5\bar{s}_6 b_{s_1,s_4}(\boldsymbol{\phi},\boldsymbol{\theta},\boldsymbol{\alpha}) b_{s_2,s_3,s_5,s_6}(\boldsymbol{\phi},\theta_2,\theta_3,2\pi-\theta_5,2\pi-\theta_6,\boldsymbol{\alpha})+b_{s_1,s_6}(\boldsymbol{\phi},\boldsymbol{\theta},\boldsymbol{\alpha}) b_{s_2,s_3,s_4,s_5}(\boldsymbol{\phi},\boldsymbol{\theta},\boldsymbol{\alpha})$ \\
         & $-\bar{s}_2b_{s_1,s_3}({\phi}_1+\frac{\pi}{2},{\phi}_3+\frac{\pi}{2},\boldsymbol{\theta},\boldsymbol{\alpha})b_{s_2,s_4,s_5,s_6}(\boldsymbol{\phi},2\pi-\theta_2,\theta_4,\theta_5,\theta_6,\boldsymbol{\alpha})$\\ &$-\bar{s}_2\bar{s}_3\bar{s}_4b_{s_1,s_5}({\phi}_1+\frac{\pi}{2},{\phi}_5+\frac{\pi}{2},\boldsymbol{\theta},\boldsymbol{\alpha})b_{s_2,s_3,s_4,s_6}(\boldsymbol{\phi},2\pi-\theta_2,2\pi-\theta_3,2\pi-\theta_4,\theta_6,\boldsymbol{\alpha})$ \\
    \hline
\end{tabular}
\caption{Recursive formula for various qubit numbers.}\label{bS}
\end{table}
\hfill
\newpage

\twocolumngrid
\begingroup\raggedright\endgroup


\begin{thebibliography}{10}

\bibitem{LIEB1961407}
E.~Lieb, T.~Schultz and D.~Mattis,
\textit{{Two soluble models of an antiferromagnetic chain}},
\href{https://doi.org/10.1016/0003-4916(61)90115-4}{Annals of Physics \textbf{16}(3), 407 (1961),}.

\bibitem{Thouless1960}
D. J. Thouless,
\textit{{Stability conditions and nuclear rotations in the Hartree-Fock theory}},
\href{https://www.sciencedirect.com/science/article/pii/0029558260900481?via%3Dihub}{Nuclear Physics \textbf{21}, 225-232 (1960)}.

\bibitem{szabo2012modern}
A. Szabo and N~S. Ostlund  2012 \textit{{ Modern quantum chemistry: introduction to
  advanced electronic structure
  theory}}\href{https://books.google.com.br/books?id=1ky8QgAACAAJ}{
  (Macmillan, 1982)}
  

\bibitem{EA2007}
P. Echenique and J. L. Alonso, \textit{{A mathematical and computa-
tional review of Hartree–Fock SCF methods in quantum chem-
istry}},
\href{https://doi.org/10.1080/00268970701757875}{Mol. Phys. \textbf{105}, 3067 (2007)},
[\href{https://arxiv.org/abs/0705.0337}{{\ttfamily arXiv:0705.0337
}}].

\bibitem{Kraus2010}
C. V. Kraus and J. I. Cirac,
\textit{{Generalized Hartree–Fock theory for interacting fermions in lattices: numerical methods}}, \href{https://iopscience.iop.org/article/10.1088/1367-2630/12/11/113004}{New Journal of Physics \textbf{12}, 113004 (2010)}, [\href{https://arxiv.org/abs/1005.5284}{{\ttfamily arXiv:1005.5284}}].


\bibitem{Terhal2023}
Y. Herasymenko, M. Stroeks, J. Helsen and B. Terhal,
\textit{{Optimizing sparse fermionic Hamiltonians}},
\href{https://quantum-journal.org/papers/q-2023-08-10-1081/}{Quantum \textbf{7}, 1081 (2023)}, [\href{https://arxiv.org/abs/2211.16518}{{\ttfamily arXiv:2211.16518}}].


\bibitem{Bravyi2017}
S. Bravyi and D. Gosset,
\textit{{Complexity of quantum impurity problems}}, 
\href{https://link.springer.com/article/10.1007/s00220-017-2976-9}{Comm. Math. Phys. \textbf{356}, 451–500 (2017)}, [\href{https://arxiv.org/abs/1609.00735}{{\ttfamily arXiv:1609.00735}}].





\bibitem{Knill2001} E. Knill, \textit{{Fermionic linear optics and matchgates}} (2001),
[\href{https://arxiv.org/abs/quant-ph/0108033}{{\ttfamily arXiv:quant-ph/0108033
}}].

\bibitem{Terhal2002}
B. M. Terhal and D. P. DiVincenzo, 
\textit{{Classical simulation of
noninteracting-fermion quantum circuit}},
\href{https://doi.org/10.1103/PhysRevA.65.032325}{, Phys. Rev. A \textbf{65},  032325 (2002)}, [\href{https://arxiv.org/abs/quant-ph/0108010}{{\ttfamily arXiv:quant-ph/0108010}}].

\bibitem{Brod2016}
D. Brod, 
\textit{{Efficient classical simulation of matchgate circuits with generalized inputs and measurements}},
\href{https://doi.org/10.1103/PhysRevA.93.062332
}{, Phys. Rev. A \textbf{93},  062332  (2016)}, [\href{https://arxiv.org/abs/1602.03539}{{\ttfamily arXiv:1602.03539
}}].

\bibitem{CEF2011}
P. Calabrese, F. H. L. Essler, M. Fagotti, 
\textit{{Quantum Quench in the Transverse Field Ising Chain}},
\href{ 	
https://doi.org/10.1103/PhysRevLett.106.227203}{Phys. Rev. Lett. \textbf{106}, 227203 (2011)}, [\href{https://arxiv.org/abs/1104.0154}{{\ttfamily  arXiv:1104.0154
}}].

\bibitem{Peschel2001}
M.-C. Chung and I.~Peschel, \textit{{Density-matrix spectra of solvable fermionic systems}},
\href{https://journals.aps.org/prb/abstract/10.1103/PhysRevB.64.064412}{Phys. Rev. B \textbf{64}, 064412 (2001)},
[\href{https://arxiv.org/abs/cond-mat/0103301}{{\ttfamily arXiv:cond-mat/0103301}}].



\bibitem{Kitaev2003}G. Vidal, J. I. Latorre, E. Rico, and A. Kitaev
\textit{{Entanglement in Quantum Critical Phenomena}}, \href{https://doi.org/10.1103/PhysRevLett.90.227902}{Phys. Rev. Lett. \textbf{90}, 227902 (2003)}, [\href{https://arxiv.org/abs/quant-ph/0211074
}{{\ttfamily arXiv:quant-ph/0211074
}}].



\bibitem{Zanardi2008}
M. Cozzini, P. Giorda, P. Zanardi,
\textit{{Quantum phase transitions and quantum fidelity in free fermion graphs}}, \href{https://journals.aps.org/prb/abstract/10.1103/PhysRevB.75.014439}{Phys. Rev. B \textbf{75}, 014439 (2007)}, [\href{https://arxiv.org/abs/quant-ph/0608059}{{\ttfamily arXiv:quant-ph/0608059}}].


\bibitem{Gluza2018}
M. Gluza, M. Kliesch, J. Eisert, and L. Aolita,
\textit{{Fidelity witnesses for fermionic quantum simulations}}, \href{https://doi.org/10.1103/PhysRevLett.120.190501}{Phys. Rev. Lett. \textbf{120}, 190501(2018)}, [\href{https://arxiv.org/abs/1703.03152
}{{\ttfamily arXiv:1703.03152
}}].


\bibitem{Gorman2022}
B. O’Gorman,
\textit{{Fermionic tomography and learning }}, [\href{https://arxiv.org/abs/2207.14787
}{{\ttfamily  arXiv:2207.14787
}}].


\bibitem{Leone2024}
L. Bittel, A. A. Mele, J. Eisert, L. Leone, 
\textit{{Optimal trace-distance bounds for free-fermionic states: Testing and improved tomography }}, [\href{https://arxiv.org/abs/2409.17953
}{{\ttfamily  arXiv:2409.17953
}}].



\bibitem{RZ2023}
 J. Zhang, M. A. Rajabpour, 
\textit{{Trace distance between fermionic Gaussian states from a truncation method  }}, \href{https://doi.org/10.1103/PhysRevA.108.022414}{Phys. Rev. A \textbf{108}, 022414(2023)}, [\href{https://arxiv.org/abs/2210.11865
}{{\ttfamily  arXiv:2210.11865
}}].


\bibitem{Wang2024}
 Q. Wang, Z. Zhang, 
\textit{{Fast Quantum Algorithms for Trace Distance Estimation }}, \href{10.1109/TIT.2023.3321121}{IEEE Transactions on Information Theory  \textbf{70}, 2720 (2024)}, [\href{https://arxiv.org/abs/2301.06783
}{{\ttfamily  arXiv:2301.06783
}}].


\bibitem{NRR2024}
 M. N. Najafi, A. Ramezanpour, M. A. Rajabpour, 
\textit{{A field theory representation of sum of powers of principal minors and physical applications }},  [\href{https://arxiv.org/abs/2403.09874
}{{\ttfamily arXiv:2403.09874
}}].

\bibitem{Surace2022}  J. Surace, and L. Tagliacozzo, \textit{Fermionic Gaussian states: an introduction to numerical approaches}, \href{https://10.21468SciPostPhysLectNotes.54}{ SciPost Phys. Lect. Notes., 054(2022) }, [\href{https://doi.org/10.48550/arXiv.2111.08343}{{\ttfamily arXiv: 2111.08343
}}].

\bibitem{Bravyi2005}
S. Bravyi,
\newblock \textit{Lagrangian representation for fermionic linear optics},
\href{https://www.rintonpress.com/journals/doi/QIC5.3-3.html}{Quantum Inf. and Comp. \textbf{5}, 216 (2005)}, [\href{https://arxiv.org/abs/quant-ph/0404180}{{\ttfamily arXiv:quant-ph/0404180}}].

\bibitem{Becca-Sorella2017}
F Becca, S Sorella,  
\newblock \textit{Quantum Monte Carlo approaches for correlated systems},  \href{https://www.cambridge.org/core/books/quantum-monte-carlo-approaches-for-correlated-systems/EB88C86BD9553A0738BDAE400D0B2900}{Cambridge University Press 2017}.


\bibitem{TKR2024}
B. Tarighi, R. Khasseh, and M. A. Rajabpour, \textit{{Efficient Representation of Gaussian Fermionic Pure States in Non-Computational
Bases}},
\href{https://journals.aps.org/pra/issues/109/6}{Phys. Rev. A \textbf{64}, 064412 (2024)},
[\href{https://arxiv.org/abs/2403.03289}{{\ttfamily arXiv:2403.03289}}].
%
\bibitem{Essler1994}
 V. E. Korepin, A. G. Izergin, F. H. L. Essler and D. B. Uglov, \textit{{Correlation function of the spin 1/2 XXX
antiferromagnet}}, 
\href{https://doi.org/10.1016/0375-9601(94)90074-4}{Phys. Lett. A \textbf{190}, 182 (1994)}, [\href{https://arxiv.org/abs/cond-mat/9403066}{{\ttfamily arXiv:cond-mat/9403066}}].



 \bibitem{Shiroishi2001}
M. Shiroishi, M. Takahashi, and Y. Nishiyama, \textit{{Emptiness formation probability for the one-dimensional
isotropic XY model}}, 
\href{https://doi.org/10.1143/JPSJ.70.3535}{J. Phys. Soc. Jpn. \textbf{70}, 3535 (2001)}, [\href{https://arxiv.org/abs/cond-mat/0106062}{{\ttfamily arXiv:cond-mat/0106062}}]

\bibitem{Franchini2005}
 F. Franchini and A. Abanov, \textit{{Asymptotics of Toeplitz determinants and the emptiness formation probability
for the XY spin chain}}, \href{https://dx.doi.org/10.1088/0305-4470/38/23/002}{J. Phys. A: Math. Gen. 38 (2005) 5069-5095}, [\href{https://arxiv.org/abs/cond-mat/0502015}{{\ttfamily arXiv:cond-mat/0502015}}]


\bibitem{Stephan2009}
 J-M. St\'ephan, S. Furukawa, G. Misguich, and V. Pasquier, \textit{Shannon and entanglement entropies of one- and two-dimensional critical wave functions},  \href{https://journals.aps.org/prb/abstract/10.1103/PhysRevB.80.184421}{Physical Review B 80, 184421 (2009)},  [\href{https://arxiv.org/abs/0906.1153}{{\ttfamily arXiv:0906.1153}}].




\bibitem{NR2016}
K.~Najafi, and M. A.~Rajabpour,
\newblock \textit{Formation probabilities and Shannon information and their time evolution after quantum quench in the transverse-field XY chain},
\href{https://journals.aps.org/prb/abstract/10.1103/PhysRevB.93.125139}{Physical Review B \textbf{93}, 125139 (2016)}, [\href{https://arxiv.org/abs/1511.06401}{{\ttfamily arXiv:1511.06401}}].

%
\bibitem{Wei2003}
T. C. Wei, P. Goldbart, \textit{{Geometric measure of entanglement and applications to bipartite and multipartite quantum states}},
\href{https://doi.org/10.1103/PhysRevA.68.042307}{Phys. Rev. A \textbf{68}, 042307 (2003)},
[\href{https://arxiv.org/abs/quant-ph/0307219}{{\ttfamily arXiv:quant-ph/0307219
}}].
%


%
\bibitem{Wei2005}
T. C. Wei, D. Das, S. Mukhopadhyay, S. Vishveshwara, and P. Goldbart, \textit{{Global entanglement and quantum criticality in spin chains}},
\href{https://journals.aps.org/pra/abstract/10.1103/PhysRevA.71.060305}{Phys. Rev. A \textbf{71}, 060305 (2005)},
[\href{https://arxiv.org/abs/quant-ph/0405162}{{\ttfamily arXiv:quant-ph/0405162}}].
%



%
\bibitem{Sen2010}
A. Sen(De) and U. Sen, \textit{{Channel capacities versus entanglement measures in multiparty quantum states}},
\href{https://doi.org/10.1103/PhysRevA.81.012308}{Phys. Rev. A \textbf{81}, 012308 (2010)},
[\href{https://arxiv.org/abs/0909.0580}{{\ttfamily  arXiv:0909.0580
}}].
%





\bibitem{Stephan2010} J-M. St\'ephan, G. Misguich, and V. Pasquier, \textit{R\'enyi entropy of a line in two-dimensional Ising model},  \href{https://journals.aps.org/prb/abstract/10.1103/PhysRevB.82.125455}{Phys.
Rev. B 82, 125455 (2010)}, [\href{https://arxiv.org/abs/1006.1605}{{\ttfamily arXiv:1006.1605}}].


\bibitem{Alcaraz2013}
F. C. Alcaraz and M. A. Rajabpour, \textit{Universal Behavior of the Shannon Mutual Information of Critical Quantum Chains}, \href{https://journals.aps.org/prl/abstract/10.1103/PhysRevLett.111.017201}{Phys. Rev. Lett. 111, 017201 (2013)}, [\href{https://arxiv.org/abs/1305.1239}{{\ttfamily arXiv:1305.1239}}].


 \bibitem{Stephan:2014}
 J.-M. St\'ephan, \textit{{Shannon and R\'enyi mutual information in quantum critical spin chains}},
  \href{https://doi.org/10.1103/PhysRevB.90.045424}{Phys. Rev. B {\bfseries
  90}, 045424 (2014)}, [\href{https://arxiv.org/abs/1403.6157}{{\ttfamily
  arXiv:1403.6157}}]. 
  
%
\bibitem{Alcaraz2014}
F. C. Alcaraz, M. A. Rajabpour,
\newblock \textit{Universal behavior of the Shannon and R{\'e}nyi mutual information of quantum critical chains},
\href{https://journals.aps.org/prb/abstract/10.1103/PhysRevB.90.075132}{Phys. Rev. B \textbf{90}, 075132 (2014)},
[\href{https://arxiv.org/abs/1405.1074}{{\ttfamily arXiv:1405.1074}}].


\bibitem{Tarighi2022} B. Tarighi, R. Khasseh, M. N. Najafi, M. A. Rajabpour, \textit{Universal logarithmic correction to Rényi (Shannon) entropy in generic systems of critical quadratic fermions}, \href{https://journals.aps.org/prb/abstract/10.1103/PhysRevB.105.245109}{Physical Review B 105 \textbf{24}, 245109 (2022)}, [\href{https://arxiv.org/abs/2203.13124}{{\ttfamily arXiv:2203.13124}}].



\bibitem{central-charge} N. U. K\"oyl\"uo\u{g}lu, S. Majumder, M. Amico, S. Mostame, E. van den Berg, M. A. Rajabpour, Z. Minev, K. Najafi, \textit{Measuring central charge on a universal quantum processor}, 
[\href{ 	
https://doi.org/10.48550/arXiv.2408.06342}{{\ttfamily arXiv:2408.06342}}].

\bibitem{Heyl2013} M. Heyl, A. Polkovnikov, and S. Kehrein, \textit{Dynamical Quantum Phase Transitions in the Transverse-Field Ising Model}, \href{https://doi.org/10.1103/PhysRevLett.110.135704}{Phys. Rev. Lett. \textbf{110}, 135704 (2013) }, [\href{https://arxiv.org/abs/1206.2505}{{\ttfamily arXiv:1206.2505
}}].

\bibitem{Heyl2018} M. Heyl, \textit{Dynamical quantum phase transitions:
a review}, \href{https://dx.doi.org/10.1088/1361-6633/aaaf9a}{Rep. Prog. Phys. \textbf{81},054001(2018) }, [\href{https://arxiv.org/abs/1709.07461}{{\ttfamily arXiv:1709.07461
}}].

\bibitem{NRV2020} K. Najafi, M. A. Rajabpour, J. Viti, \textit{Return amplitude after a quantum quench in the XY chain}, \href{https://dx.doi.org/10.1088/1742-5468/ab3413}{J. Stat. Mech.  083102,(2019) }, [\href{https://arxiv.org/abs/1905.01272}{{\ttfamily arXiv:1905.01272
}}].

\bibitem{Jurcevic2017} P. Jurcevic, H. Shen, P. Hauke, C. Maier, T. Brydges, C. Hempel, B. P. Lanyon, M. Heyl, R. Blatt, and C. F. Roos, \textit{Direct Observation of Dynamical Quantum Phase Transitions in an Interacting Many-Body System}, \href{https://doi.org/10.1103/PhysRevLett.119.080501}{Phys. Rev. Lett.  \textbf{119},  080501(2017) }, [\href{https://arxiv.org/abs/1612.06902}{{\ttfamily arXiv:1612.06902
}}].

\bibitem{Najafi2021}
A. M. Gomez, S. F. Yelin, K. Najafi,
\newblock \textit{Born Machines for Periodic and Open XY Quantum Spin Chains},
[\href{https://arxiv.org/abs/2112.05326}{{\ttfamily arXiv:2112.05326}}].
%
\bibitem{ZR2025}  J. Zhang, M. A. Rajabpour, \textit{Pfaffian Decomposition Ansatz for quantum tomography of Ising systems}, {\text{in preparation}}


\bibitem{R2015}  M. A. Rajabpour, \textit{Post measurement bipartite entanglement entropy in conformal field theories }, \href{https://doi.org/10.1103/PhysRevB.92.075108}{Phys. Rev. B.  \textbf{92},  075108(2015) }, [\href{https://arxiv.org/abs/1501.07831}{{\ttfamily arXiv:1501.07831
}}].

\bibitem{Baweja2024}  K. Baweja, D. J. Luitz, and S. J. Garratt, \textit{Post-measurement Quantum Monte Carlo},[\href{https://doi.org/10.48550/arXiv.2410.13844}{{\ttfamily arXiv:2410.13844
}}]

\bibitem{Potter2024}  Z. Cheng, R. Wen, S. Gopalakrishnan, R. Vasseur, A. C. Potter, \textit{Universal structure of measurement-induced information in many-body ground states }, \href{https://doi.org/10.1103/PhysRevB.109.195128}{Phys. Rev. B.  \textbf{109}, 195128(2024) }, [\href{https://arxiv.org/abs/2312.11615}{{\ttfamily arXiv:2312.11615
}}].


\bibitem{Lieb2016}
A. Giuliani, I. Jauslin,  E. H. Lieb,  \textit{{A Pfaffian Formula for Monomer–Dimer Partition Functions}}, \href{https://doi.org/10.1007/s10955-016-1484-1}{J Stat Phys 163, 211–238 (2016)},

\bibitem{Lieb1968}
E. H. Lieb, \textit{{A theorem on Pfaffians}}, \href{https://doi.org/10.1016/S0021-9800(68)80078-X}{J. Comb. Theory 5, 313–319 (1968)}.

\bibitem{Rajabpour2015}  M. A. Rajabpour, \textit{Post-measurement bipartite entanglement entropy in conformal field theories}, \href{https://doi.org/10.1103/PhysRevB.92.075108}{ Phys. Rev. B.  \textbf{92},  075108(2015) }, [\href{https://doi.org/10.48550/arXiv.1501.07831}{{\ttfamily arXiv:1501.07831
}}]

\bibitem{Numasawa2016} T. Numasawa, N. Shiba, T. Takayanagi, and K. Watanabe,\textit{EPR pairs, local projections and quantum teleportation in holography}, \href{https://doi.org/10.1007/JHEP08(2016)077}{JHEP. \textbf{2016},  1--52(2016) }, [\href{https://doi.org/10.48550/arXiv.1604.01772}{{\ttfamily arXiv:1604.01772
}}]

\bibitem{Guo2019}  W. Z. Guo, \textit{Entanglement of purification and projective measurement in CFT}, \href{https://doi.org/10.1016/j.physletb.2019.134934}{Phys. Lett. B. \textbf{797},  134934(2019) }, [\href{https://doi.org/10.48550/arXiv.1901.00330}{{\ttfamily arXiv:1901.00330
}}]
\bibitem{Antonini2022}  S. Antonini, G. Bentsen, C. Cao, J. Harper, S. K. Jian, and B. Swingle, \textit{Holographic measurement and bulk teleportation}, \href{https://doi.org/10.1007/JHEP12(2022)124}{JHEP.  \textbf{2022},  1--76(2022) }, [\href{ 	https://doi.org/10.48550/arXiv.2209.12903}{{\ttfamily arXiv:2209.12903
}}]

\bibitem{Antonini2023} S. Antonini, B. Grado-White, S. K. Jian, and B. Swingle,\textit{Holographic measurement in CFT thermofield doubles}, \href{https://doi.org/10.1007/JHEP07(2023)014}{JHEP. \textbf{2023},  1--56(2023) }, [\href{https://doi.org/10.48550/arXiv.2304.06743}{{\ttfamily arXiv:2304.06743
}}]

\bibitem{Weinstein2023}  Z. Weinstein, R. Sajith, E. Altman,  and S. J. Garratt, \textit{Nonlocality and entanglement in measured critical quantum Ising chains}, \href{https://10.1103/PhysRevB.107.245132}{ Phys.
Rev. B.  \textbf{107},  245132(2023) }, [\href{https://doi.org/10.48550/arXiv.2301.08268}{{\ttfamily arXiv:2301.08268
}}]

\bibitem{R2016}  K. Najafi, M. A. Rajabpour, \textit{Entanglement entropy after selective measurements in quantum chains}, \href{https://doi.org/10.1007/JHEP12(2016)124}{JHEP.  \textbf{2016},  1--73(2016) }, [\href{https://doi.org/10.48550/arXiv.1608.04074}{{\ttfamily arXiv:1608.04074
}}]



\bibitem{Mele2025}  A. A. Mele, and Y. Herasymenko, \textit{Efficient learning of quantum states prepared with few fermionic non-Gaussian gates}, \href{https://10.1103/PRXQuantum.6.010319}{ Phys.
Rev. X.  \textbf{6},  010319(2025) }, [\href{https://doi.org/10.48550/arXiv.2402.18665}{{\ttfamily arXiv:2402.18665
}}]
\end{thebibliography}
\end{document}